\documentclass[twocolumn]{autart}    

 \usepackage[T1]{fontenc}    
 \usepackage{amsfonts}       
 \usepackage{amsmath} 
 \usepackage{amssymb}  
 \usepackage{mathrsfs}
 \usepackage{algorithm}
 \usepackage{algpseudocode}
 \algnewcommand{\Inputs}[1]{%
  \State \textbf{Inputs:}
  \Statex \hspace*{\algorithmicindent}\parbox[t]{.8\linewidth}{\raggedright #1}
}
\algnewcommand{\Initialize}[1]{%
  \State \textbf{Initialize:}
  \Statex \hspace*{\algorithmicindent}\parbox[t]{.8\linewidth}{\raggedright #1}
}

\usepackage[a4paper]{geometry}
 \usepackage{bm}
 \usepackage{array}
 \usepackage{stfloats}
 \usepackage{graphicx}
 \usepackage[colorlinks=true, allcolors=black]{hyperref}
 \DeclareMathAlphabet{\mathcal}{OMS}{cmsy}{m}{n}
 \DeclareSymbolFont{largesymbols}{OMX}{cmex}{m}{n}
 \usepackage[normalem]{ulem}
 
\newcommand{\mA}{\mathcal{A}}

\newcommand{\R}{\mathbb{R}}
\newcommand{\mbE}{\mathbb{E}}

\newcommand{\mO}{\mathcal{O}}
\newcommand{\tmO}{\tilde{\mathcal{O}}}
\newcommand{\tx}{\tilde{x}}
\newcommand{\tu}{\tilde{u}}
\newcommand{\ty}{\tilde{y}}

\newcommand{\hPhi}{\hat{\Phi}}

\newcommand{\hx}{\hat{x}}
\newcommand{\hy}{\hat{y}}
\newcommand{\hv}{\hat{v}}
\newcommand{\hH}{\hat{H}}
\newcommand{\hP}{\hat{P}}
\newcommand{\hbw}{\hat{\bm{w}}}
\newcommand{\bbw}{\bar{\bm{w}}}
\newcommand{\tbw}{\tilde{\bm{w}}}

\newcommand{\bw}{\bm{w}}
\newcommand{\bW}{\bm{W}}

\newcommand{\bu}{\bm{u}}
\newcommand{\bM}{\bm{M}}
\newcommand{\accw}{\bm{w}^{o}}

\newcommand{\starM}{\bm{M}^{*}}

\newcommand{\defeq}{\overset{i.e.}{=}}

\newcommand{\tr}[1]{\text{tr}(#1)}

\newcommand{\dimm}[1]{\dim{(\mathbb{#1})}}
\newcommand{\proj}[2]{\text{Proj}_{\mathbb{#1}}\bigl(#2\bigr)}
\newcommand{\pto}[1]{\|#1\|_{\psi_2}}
\newcommand{\rgt}{\texttt{rgt}}
\newcommand{\hf}{\frac{1}{2}}

\newtheorem{definition}{Definition}[section]
\newtheorem{lemma}{Lemma}[section]
\newtheorem{assumption}{Assumption}
\newtheorem{remark}{Remark}[section]

\newtheorem{corollary}{Corollary}

\begin{document}

\begin{frontmatter}

\title{Data-Driven Adversarial Online Control\\ for Unknown Linear Systems\thanksref{footnoteinfo}}
\thanks[footnoteinfo]{This paper was not presented at any IFAC 
meeting.}

\author[Atlanta]{Zishun Liu}\ead{zliu910@gatech.edu}, 
\author[Atlanta]{Yongxin Chen}\ead{yongchen@gatech.edu}              

\address[Atlanta]{Department of Aerospace Engineering, Georgia Institute of Technology, Atlanta, GA 30332, USA}  


\begin{abstract}                          
We consider the online control problem with an unknown linear dynamical system in the presence of adversarial perturbations and adversarial convex loss functions. Although the problem is widely studied in model-based control, it remains unclear whether data-driven approaches, which bypass the system identification step, can solve the problem. In this work, we present a novel data-driven online adaptive control algorithm to address this online control problem. Our algorithm leverages the behavioral systems theory to learn a non-parametric system representation and then adopts a perturbation-based controller updated by online gradient descent. We prove that our algorithm guarantees an $\tmO(T^{2/3})$ regret bound with high probability, which matches the best-known regret bound for this problem. Furthermore, we extend our algorithm and performance guarantee to the cases with output feedback.
\end{abstract}

\end{frontmatter}

\section{Introduction}
\label{sec:introduction}
In recent years, the popularity of data-driven control approaches has surged. Unlike model-based methods which hinge on the accurate identification of system parameters, algorithms for data-driven control bypass the system identification step. Instead, they leverage the Fundamental Lemma in Behavioral System Theory \cite{2005Anote} to construct a non-parametric system representation based directly on data \cite{hou2013model}. In applications where the system is complex and process data are readily available, data-driven control method stands out for its simplicity, generality, and robustness, offering a viable and practical alternative to the conventional model-based approach \cite{markovsky2021behavioral}. By now, data-driven scheme has already been successfully implemented in a variety of control problems, e.g., system stabilization \cite{de2019formulas,de2021low}, LQR \cite{8453019} and model predictive control \cite{8795639,berberich2020data}. 

Despite the successes of data-driven control in various domains, one of the significant challenges that persist is the problem of adversarial online control, whose significance has grown alongside the flourishing of online learning in adversarial environments. It aims to design a control policy to address the following problem
\begin{subequations}\label{problem: intro}
    \begin{align}
        &\min J=\sum_{t=0}^{T-1}c_t(u_t,x_t)\\
        \text{s.t.}\quad &x_{t+1}=g_t(x_t,u_t)+w_t
    \end{align}            
\end{subequations}
where $c_t$ is the instantaneous cost function, $x_t$ is the state, $u_t$ is the control and $w_t$ is the disturbance. At each time step $t$, the learner makes a decision based on its observations and historical data, and subsequently receives the next state $x_{t+1}$ along with the instantaneous cost $c_t$. Unlike standard stochastic control, both $c_t$ and $w_t$ in online control can be given in an adversarial form and there is no assumption on their statistical properties, making it impossible to find the optimal control policy by Riccati Equation or HJB Rule {\em a priori}. Under this circumstance, the performance of an adversarial online control algorithm is typically evaluated by the cost gap between the algorithm and a specified policy. This evaluation can be either asymptotic, considering an infinite number of time steps, or non-asymptotic, focusing on a finite number of time steps.

Research efforts in adversarial online control span a wide array of scenarios, addressing challenges in known time-varying systems \cite{agarwal2019online}, unknown time-invariant systems \cite{hazan2020nonstochastic,chen2021black}, and nonlinear systems \cite{ho2021online,li2023online}. However, the algorithms proposed in these works rely either on a pre-known system model or an estimation of the parameterized model. Concerning data-driven approaches, although there have been works in handling data-driven stochastic online control problems, e.g., \cite{9028943,bianchin2023online}, existing studies on adversarial online control with data-driven approaches are scarce. The work \cite{berberich2020data} delves into online control problems with output feedback, unknown system models, and adversarial measurement noises, and proposes a data-driven model predictive control scheme. While this approach ensures the stability of the system, it falls short in providing insights into its performance. In \cite{de2021low} the authors investigate the optimal control problem with adversarial disturbance and propose a data-driven algorithm based on semi-definite programming (SDP), but it only guarantees a bounded asymptotic performance, which corresponds to a linear non-asymptotic regret bound. The absence of online updates to the controller leads to a lack of guarantees on achieving sublinear non-asymptotic regret bounds. To this end, the question of whether methods in a data-driven manner are capable of tackling adversarial online control as well as model-based methodologies naturally arises.

\textit{For adversarial online control problem with an unknown dynamical system, can we develop an algorithm that 1) is in data-driven manner, 2) guarantees a sublinear regret bound?}

In this work, we address the adversarial online control problem over a linear time-invariant (LTI) system. The contributions of our algorithm are as follows. First, instead of identifying the system matrices $(A, B)$, our algorithm leverages techniques in Statistics and Behavioral Systems Theory to learn a non-parametric representation of the system. To the best of the authors' knowledge, this is the first work that bridges the fundamental lemma and adversarial online control algorithm. Second, the algorithm refers to the accumulated disturbance-action controller (ADAC) and adopts an adaptive ADAC to generate an input sequence. Compared to the disturbance-action controller (DAC) in model-based adversarial online control, our method is more suitable for a data-driven framework and can be extended to the output feedback case directly. Third, in theoretical analysis, we prove that under mild assumptions, our algorithm guarantees an $\tmO(T^{2/3})$ regret bound with high probability, which matches the best-known result achieved by model-based methods. Moreover, we show that just changing state sequence $\{x_t\}$ to output sequence $\{y_t\}$ and our algorithm can be directly applied to the output feedback case while keeping the same $\tmO(T^{2/3})$ regret bound.

The rest of the paper is organized as follows. Section \ref{sec: bg} formulates the adversarial online control problem and provides background on behavioral systems theory and perturbation-based controllers. Section \ref{sec: previous} focuses on a special case of the problem we study.
We present the main algorithm of this paper in Section \ref{sec: DD-NOC} and the regret analysis in Section \ref{sec: analysis}. In Section \ref{sec: output} we extend our method to systems with output feedback. This is followed by a concluding remark in Section \ref{sec: conclu}.

\textit{Notations.} For an integer $a$, $[a]=\min\{n\in\mathbb{N}: n\geq a\}$. For a vector $x\in\R^n$, $\|x\|$ is its Euclidean norm. For a matrix $A=\{a_{ij}\}_{m\times n}$, $a_{ij}$ denotes its $(i,j)$-th element, $A^T$ denotes its transpose, $\tr{A}$ denotes its trace if $A$ is square, $\|A\|$ denotes its operator norm, and $M_t^{(a:b)}=\{M_t^{(a)},\dots,M_t^{(b)}\}$. For a bounded set $\mathbb{X}$, $\dimm{X}=\max_{x,y\in\mathbb{X}}{\|x-y\|}$. For a sequence  $x=\{x_k\}_{k=1}^{N-1}$, $x_{a:b}$ denotes $\{x_a,\cdots,x_b\}$ listed in a column, and we define its Hankel matrix with length $L$ as 
\begin{equation}\label{eq: Hankel}
	H_L(x)=\begin{bmatrix}
		x_0 & x_1 & \cdots & x_{N-L}\\
		x_1 & x_2 & \cdots & x_{N-L+1}\\
		\vdots & \vdots & \ddots & \vdots\\
		x_{L-1} & x_L & \cdots & x_{N-1}
	\end{bmatrix}
\end{equation}
and $H_L(x)[k,:]=\begin{bmatrix}
		x_{k-1} & x_k & \cdots & x_{N-L+k-1}
	\end{bmatrix}$.
We adopt $\mO$ notation for complexity and $\tmO(\cdot)$ to omit $\log$ terms. We use $\mathcal{S}^{n-1}$ to denote the unit sphere on $\R^n$ and $\text{Unif}(\mathcal{S}^{n-1})$ to denote the uniform distribution on it. 


\section{Problem Formulation and Preliminaries}
\label{sec: bg}
\subsection{Problem Formulation}
Consider the online control problem over an LTI system
\begin{subequations}\label{sys: noise-LTI}
    \begin{align}
        &\min J=\sum_{t=0}^{T-1}c_t(u_t,x_t)\\
        \text{s.t.}\quad &x_{t+1}=Ax_t+Bu_t+w_t,
    \end{align}            
\end{subequations}
where $x_t\in\mathbb{R}^n$, $u_t\in\mathbb{R}^m$, and $w_t\in\mathbb{R}^n$ is a disturbance. Here the true dynamics $A\in \mathbb{R}^{n\times n}$ and $B\in \mathbb{R}^{n\times m}$ are unknown and we have no prior knowledge of how $w_t$ is generated other than its boundness. To permit a discussion on non-asymptotic regret bounds, we assume $(A, B)$ is controllable and $A$ is $(\kappa,\rho)$-exponentially stable (see e.g., \cite{li2023online}), as defined below.

\begin{definition}
    System \eqref{sys: noise-LTI} is said to be $(\kappa,\rho)$-exponentially stable if $\|A^k\|\leq \kappa\rho^k$ with some $\kappa>0$, $0<\rho<1$ and $\forall k\in \mathbb{N}_+$.
\end{definition}

\begin{assumption} \label{as: AB}
The pair $(A,B)$ is controllable and $A$ is $(1,\rho)$-exponentially stable with some $0<\rho<1$.
\end{assumption}

\begin{assumption} \label{as: bounded_w}
$w_t$ is bounded by $\|w_t\|\leq\varepsilon$.
\end{assumption}

At every time step $t$, the learner plays $u_t$ and then observes $x_{t+1}$ and instantaneous cost $c_t(u_t,x_t): \mathbb{R}^{m+n}\to [0,\infty)$. It should be noted that $c_t$ is revealed only after playing $u_t$, meaning it could be given in an adversarial form.

\begin{assumption} \label{as: cost}
At any time $t$, $c_t(u_t,x_t)$ is a convex and differentiable function of $x_t$ and $u_t$, and $\|\nabla_{u_t,x_t}c_t\|\leq G$ with some finite constant $G>0$.
\end{assumption}

 Since we have no assumption on the statistical properties of $w_t$, it is impossible to chase the ``minimal'' total cost $\sum_{t=0}^{T-1}c_t(u_t,x_t)$ over a finite timeline $T$ \cite{hazan2022introduction}. However, if we fix a reference policy class $\Pi$ and suppose that there exists an oracle that knows everything {\em a priori} and can choose the best policy $\pi^{*}=\arg\min_{\pi\in\Pi}\sum_{t=0}^{T-1}{c_t(u_t^\pi,x_t^\pi)}$ based on prior knowledge, then we can use \textit{policy regret} to measure how good a policy is compared to $\pi^{*}$. The policy regret is defined as follows, and $\Pi$ used throughout this paper will be given in Section \ref{subsec: ADAC}.

\begin{definition}\label{def: regret}
Given a policy class $\Pi$, the policy regret between the learner's policy $\mathcal{A}$ and $\Pi$ over $T$ steps is defined as
\begin{equation}
    \verb|rgt|_T(\mathcal{A},\Pi)=\sum_{t=0}^{T-1}{c_t(u_t,x_t)}-\min_{\pi\in\Pi}{\sum_{t=0}^{T-1}{c_t(u_t^\pi,x_t^\pi)}}
\end{equation}
where $x_t$ is the actual state, $u_t$ is generated by $\mathcal{A}$, and $(u_t^\pi,x_t^\pi)$ are the artificial state sequence and controls under the policy $\pi$, i.e., $x_{t+1}^\pi=Ax_t^\pi+Bu_t^\pi+w_t$, $x_0^\pi=x_0$, $u_t^\pi = \pi(x_t^\pi)$. 
\end{definition}

Our goal is to design a data-driven algorithm that can achieve a sub-linear regret, i.e., $\verb|rgt|_{T}(\mathcal{A},\Pi)\leq\tilde{\mathcal{O}}(T^\alpha)$ with some $\alpha<1$ for a reasonable $\Pi$. Before presenting our method, we 
introduce the data-driven representation.

\subsection{Non-Parametric Representation of LTI Systems}
Consider the setting where we can acquire a noise-free trajectory of \eqref{sys: noise-LTI} with $w_t=0$. In this case, by making use of Fundamental Lemma, we can give a precise non-parametric representation of the LTI system, which is stated in the following theorem.

\begin{thm}[\cite{berberich2020trajectory}, Theorem 3]\label{thm: Hmatrix}
Suppose $\{u,x\}$ is a noise-free trajectory of the system (\ref{sys: noise-LTI}), where $\{u\}$ is persistently exciting of order $L+2n$. Then, $\{\hat{u},\hat{x}\}=\{(\hat{u}_1,\hat{x}_1),\dots,(\hat{u}_L,\hat{x}_L)\}$ is a noise-free trajectory of system (\ref{sys: noise-LTI}) if and only if there exists $\alpha\in\mathbb{R}^{N-L+1}$ such that
\begin{equation} \label{eq: DD-Represent}
    \begin{bmatrix}
        H_L(u) \\ H_L(x)
    \end{bmatrix}\alpha=\begin{bmatrix}
        \hat{u} \\ \hat{x}
    \end{bmatrix}.
\end{equation}
\end{thm}

The above theorem indicates that if we can obtain a noise-free trajectory $\{x,u\}$ of an LTI system, then the Hankel matrices $H_L(u), H_L(x)$ give a precise non-parametric representation of the system behaviors. Throughout this paper, we use this representation of LTI systems for control tasks. 


\section{Simplified Setting: Control with A Clean Trajectory}\label{sec: previous}
We begin with a simplified setting where we assume access to a clean trajectory  $\{u^d,x^d\}=\{u_t^d,x_t^d\}_{t=0}^{N-1}$ such that $x_{t+1}^d=Ax_t^d+Bu_t^d$. The results will be extended to general settings via an Explore-Then-Commit (ETC) strategy in Section \ref{sec: DD-NOC}.
An algorithm using ETC first explores system behavior randomly, meanwhile estimates the system by exploration, and then treats the estimated system as true and conducts decision-making methods on it. In the adversarial online control problem, if we can find an efficient data-driven algorithm $\mA_0$ which invokes $\{u^d,x^d\}$ and guarantees sublinear regret, then it becomes possible to design an algorithm $\mA$ which first estimates a clean trajectory $\{u^d,\hx^d\}$ from perturbed data and then deploy $\mA_0$ while substituting $\{u^d,x^d\}$ for $\{u^d,\hx^d\}$ to achieve a sublinear regret. 

In the following, we discuss the reference policy class considered in this paper, including the concept of ``accumulated disturbance'', and propose an algorithm in a data-driven manner that requires a clean trajectory as input and guarantees $\tmO(\sqrt{T})$ regret bound. 

\subsection{Accumulated Disturbance and ADAC} \label{subsec: ADAC}
Before designing an adversarial online control algorithm, one needs to fix a reference policy class. While the linear feedback controller (LFC) $u_t=Kx_t$ is a common choice in many control problems, it is not suitable for online control problems \eqref{sys: noise-LTI} due to the non-convexity of $J(K)=\sum_{t=0}^{T-1}c_t(u_t(K),x_t(K))$ with respect to $K$ \cite[Lemma 2]{fazel2018global}. 

A more suitable alternative to the LFC is the perturbation-based controller, which generates control input based on the historical disturbance. For model-based methods, a widely used perturbation-based controller is the disturbance action controller (DAC) proposed in \cite{agarwal2019online} defined as follows.
\begin{definition} \label{def: dac}
Given the LTI system (\ref{sys: noise-LTI}) and one of its stabilizing $K$, i.e., $A+BK$ is $(\kappa,\rho)$-exponentially stable, if we have historical access to $w_\tau$, $\tau\leq t$ at each time $t$, then Disturbance Action Controller (DAC) parameterized by $\bM=[M^{(1)},\dots,M^{(L)}], M^{(i)}\in \R^{m\times n}, i=1,\dots,L$ and $K$ is defined as
\begin{equation}\label{DAC}
    u_t^{(DAC)}=Kx_t+\sum_{i=1}^L M^{(i)}w_{t-i}
\end{equation}
\end{definition}
It turns out that if $c_t(u_t,x_t)$ is convex for all $t$, and $u_t$ is generated by DAC, then the function $J(\bM)=\sum_{t=0}^{T-1}c_t(u_t(\bM),x_t(\bM))$ is a convex function with respect to $\bM$ \cite{hazan2020nonstochastic}. 

In the data-driven problem context, however, $w_t$ is not accessible as the value of $(A,B)$ is unknown. Despite the inaccessibility of $w_t$, one with a clean trajectory $\{u^d,x^d\}$ can get access to the “accumulation” of disturbance by leveraging the properties of Hankel matrices $H_L(u^d), H_L(x^d)$. To begin with, let us give a formal definition of accumulated disturbance.
\begin{definition}\label{def: accNoise}
    Given the LTI system (\ref{sys: noise-LTI}), the accumulated disturbance $\bw_t$ is defined as 
\begin{equation}\label{eq: def of acc_noise}
    \bw_t=\sum_{i=0}^{t}{A^iw_{t-i}}.
\end{equation}
\end{definition}
For the system \eqref{sys: noise-LTI} with $x_0=0$, we have
\begin{equation}\label{eq: x_t=}
    x_t=\sum_{i=1}^t{A^{i-1}Bu_{t-i}}+\sum_{i=0}^{t-1}{A^iw_{t-1-i}}.
\end{equation}
Define $x_t^{c}=\sum_{i=1}^t{A^{i-1}Bu_{t-i}}$, then  
$x_t^{c}$ satisfies 
\begin{equation}\label{eq: x_t=xtz+bw}
    x_t=x_t^{c}+\bw_{t-1},
\end{equation}
and
\begin{equation}\label{eq: x^{zn}_t=}
    x^{c}_t=A x_{t-1}^c+Bu_{t-1},\quad x_0^z=x_0=0.
\end{equation}
Equation \eqref{eq: x^{zn}_t=} indicates $\{x_\tau^c,u_\tau\}_{\tau\leq t}$ is a clean trajectory of the system $(A,B)$. Since $\bw_{t-1}$ can be regarded as the \textit{accumulation} of disturbance $w_\tau$, $\tau=1, \ldots, t-1$ on a clean state $x_t^{z}$, we termed $\bw_t$ the \textit{accumulated disturbance}.

Next is about how to calculate $\bw_t$. Since both $\{u_t,x_t^c\}$ in \eqref{eq: x_t=xtz+bw} and $\{x^d,u^d\}$ are clean trajectories of \eqref{sys: noise-LTI}, by Theorem \ref{thm: Hmatrix}, we know there exists an $\alpha\in\mathbb{R}^{N-L+1}$ such that
\begin{equation} \label{eq: u;x'=H;H}
    \begin{bmatrix}
        u_{t-L+1:t} \\ x_{t-L+1:t}-\bw_{t-L:t-1}
    \end{bmatrix}=\begin{bmatrix}
        H_L(u^d) \\ H_L(x^d)
    \end{bmatrix}\alpha
\end{equation}

When $H_L(u^d)$ is full row rank, then every $\alpha$ that satisfies \eqref{eq: u;x'=H;H} also satisfies
\begin{equation} \label{eq: u;x'[1]=H;H[1]}
    \begin{bmatrix}
        u_{t-L+1:t} \\ x_{t-L+1}-\bw_{t-L}
    \end{bmatrix}=\begin{bmatrix}
        H_L(u^d) \\ H_L(x^d)[1,:]
    \end{bmatrix}\alpha,
\end{equation}
and each row of $x_{t-L+2:t}-\bw_{t-L+1:t-1}$ can be represented by the elements of \eqref{eq: u;x'[1]=H;H[1]}. This discovery provides a way to calculate $\bw_t$ in a data-driven manner, which is stated in the following lemma.

\begin{lemma}\label{lemma: previous work}
Suppose Assumption \ref{as: AB}, \ref{as: bounded_w} hold. Then, given a clean trajectory $\{x^d,u^d\}=\{x_1^d,u_1^d,\dots,$ $x_{N}^d,u_{N}^d\}$ of \eqref{sys: noise-LTI} where $u^d$ is persistently exciting of order $L+2n$, the sequence $\{x_0,u_0,\dots,x_{L-1},u_{L-1}\}$ is a trajectory of \eqref{sys: noise-LTI} if and only if there exists an $\alpha\in\mathbb{R}^{N+L-1}$ such that
    \begin{equation}\label{eq: lemma1 eq1}
        \begin{bmatrix}
			u_{1:L} \\ x_{1}-\bw_{0}
		\end{bmatrix}=\begin{bmatrix}
			H_L({u^{d}}) \\ H_L(x^d)[1,:]
		\end{bmatrix}\alpha.
    \end{equation}
Moreover, in this case,
    \begin{equation}\label{eq: lemma1 eq2}
        \bw_{i-1}=x_i-H_L(x^d)[i,:]\alpha, ~~ 1\leq i\leq L.
    \end{equation}
\end{lemma}

Based on Lemma \ref{lemma: previous work}, we present a pseudo-code, AccNoise(), for computing $\bw_t$, as depicted in Algorithm \ref{alg: AccNoise}. By running this program, 
\begin{equation}\label{eq: call AccNoise}
\begin{split}
    \bw_t=\text{AccNoise}(&u_{t-L+2:t}, x_{t-L+2},x_{t+1}, \\
    &\bw_{t-L+1}, H_L(u^{d}),H_L(x^d)).
\end{split}
\end{equation}

With access to $\bw_t$, we can define a class of controller named \textit{Accumulated Disturbance Action Controller} in a similar manner to DAC. An assumption is given hereafter to make gradient-based methods feasible.

\begin{definition} \label{def: adac}
Given the LTI system (\ref{sys: noise-LTI}), one of its stabilizing $K$, a set $\mathbb{M}$, and accumulated disturbance $\bw_\tau$, $\tau<t$ at each time $t$, then Accumulated Disturbance Action Controller (ADAC) parameterized by $\bM=[M^{(1)},\dots,M^{(L)}]\in\mathbb{M}$ and $K$ is
	\begin{equation}\label{eq: ADAC}
		u_t^{(ADAC)}=Kx_t+\sum_{i=1}^L M^{(i)}\bw_{t-i}
	\end{equation}
\end{definition}

\begin{assumption}\label{as: set M}
    $\mathbb{M}$ is a convex set and $\|M^{(i)}\|\leq D$ for any $\bM=[M^{(1)},\dots,M^{(L)}]\in\mathbb{M}$.
\end{assumption}

Since $u_t^{(ADAC)}$ is composed of a stabilizing state feedback and a bounded input, the LTI system \eqref{sys: noise-LTI} remains stable while applying ADAC. Besides, $J(\bM)=\sum_{t=0}^{T-1}c_t(\tu_t(\bM),\tx_t(\bM))$ is a convex function w.r.t $M$ since $u_t^{(ADAC)}$ is a linear combination of past accumulated disturbances. 
Moreover, \cite{foster2020logarithmic} shows that if the cost function is quadratic and stationary, then the optimal controller is exactly in the form of ADAC. For these reasons, we choose ADAC as the reference policy class of the problem we study, and specify the regret as
\begin{equation*}
\begin{split}
    &\rgt_T(\mA,\Pi) \\
    &=\sum_{t=0}^{T-1}c_t(u_t,x_t)-\sum_{t=0}^{T-1}c_t(u_t(\starM|\{\bw_t\}),x_t(\starM|\{\bw_t\}))
\end{split}
\end{equation*}
where 
\begin{equation}\label{eq: def of M*}
    \begin{cases}
    \starM=\arg\min_{\bM\in\mathbb{M}}\sum_{t=0}^{T-1}{c_t(u_t(\bM|\{\bw_t\}),x_t(\bM|\{\bw_t\}))} \\
    u_t(\bM|\{\bw_t\})=\sum_{i=1}^L M^{(i)}\bw_{t-i} \\
    x_{t+1}(\bM|\{\bw_t\})=Ax_{t}(\bM|\{\bw_t\})+Bu_t(\bM|\{\bw_t\})+w_t
\end{cases}
\end{equation}

\subsection{Algorithm with Clean-Trajectory Input}
To reduce the regret, we develop an adaptive ADAC policy whose parameter $\bM$ is updated by online gradient descent (OGD) as 
\begin{equation}\label{eq: previous OGD}
    \begin{split}
        u_t & =\sum_{i=1}^L M_t^{(i)}\bw_{t-i} \\
        f_t(\bM_t)&=c_t(u_t(\bM_t|\{\bw_t\}),x_t(\bM_t|\{\bw_t\})) \\
        \bM_{t+1} &=\proj{M}{\bM_t-\lambda\nabla f_t(\bM_t)}
    \end{split}
\end{equation}
It can be seen that the value of $x_t(\bM_t|\{\bw_t\})$ is necessary for operating OGD. Fortunately, given $\{\bw_\tau\},\tau\leq t$, one can simulate $x_\tau(\bM_t|\{\bw_t\})$ for any $\tau\leq t$ by leveraging Lemma \ref{lemma: previous work}, just letting $u_t=\sum_{i=1}^L M_t^{(i)}w'_{t-i}$ in \eqref{eq: lemma1 eq1} and computing \eqref{eq: lemma1 eq2}. The simulation program PiTraj() is organized in Algorithm \ref{alg: PiTraj}. One with clean trajectory $\{x^d,u^d\}$ is able to recover $x_t(\bM_t|\{\bw_t\})$ by 
\begin{equation}\label{eq: call PiTraj}
    x_t(\bM_t|\{\bw_t\})=\text{PiTraj}(\bw_{0:t-1}, \bM_t, H_L(u^{d}), H_L(x^d), t).
\end{equation}

Our algorithm for adversarial control with a clean trajectory is summarized in Algorithm \ref{alg: noise-free}. At every time step $t$, the algorithm plays $u_t =\sum_{i=1}^L M_t^{(i)}w'_{t-i}$, receives $x_{t+1}$ and $c_t$ thereafter, then calculates $\bw_t$ and simulates $x_t(\bM_t|\{\bw_t\})$ respectively, and finally updates $\bM_t$ by OGD.

\subsection{Regret Guarantees}
The following result shows that Algorithm \ref{alg: noise-free} guarantees an $\tmO(\sqrt{T})$ regret bound in the scenario where a clean trajectory is known. 
\begin{thm}\label{thm: sqrt-T regret}
Suppose Assumption \ref{as: AB},\ref{as: bounded_w},\ref{as: cost} hold and $u^d$ is persistently exciting of order $L+2n$. Then Algorithm 1 denoted as $\mathcal{A}_0$ guarantees that
\begin{equation*}
    \verb|rgt|_{T}(\mathcal{A}_0,\Pi)\leq C_0\sqrt{T}
\end{equation*}
with some constant $C_0=poly(G,\|B\|,m,n,\rho,L)$.
\end{thm}
We omit the proof of Theorem \ref{thm: sqrt-T regret} as it is a simplified version of Theorem \ref{thm: final result} in Section \ref{sec: analysis}. This $\tmO(\sqrt{T})$ regret bound is on par with model-based adversarial online control methods with a known system model. 

 \begin{algorithm}
	\caption{AccNoise $\bigl(u_{t-L+2:t}$, $x_{t-L+2}$, $x_{t+1}$, $\bw_{t-L+1}$, $H_L(u^d)$, $H_L(x^d)\bigr)$}
    \label{alg: AccNoise}
	\begin{algorithmic}[1]
	\State Find an $\alpha$ such that 
 \begin{equation*} \label{eq: AccNoise}
		\begin{bmatrix}
			\bu \\ x_{t-L+2}-\bw_{t-L+1}
		\end{bmatrix}=\begin{bmatrix}
			H_L(u^d) \\ H_L(x^d)[1,:]
		\end{bmatrix}\alpha
	\end{equation*}
	\State Calculate $\bw_t=x_{t+1}-H_L(x^d)[L,:]\alpha$. \\
	\Return{$\bw_t$}.
 \end{algorithmic}
\end{algorithm}

 \begin{algorithm}
	\caption{PiTraj $\bigl(\bw_{0:t-1}$, $\bM_t$, $H_L(u^d)$, $H_L(x^d)$, $t\bigr)$}
	\label{alg: PiTraj}
	\begin{algorithmic}[1]
     \State Initialization: $\tx_{t\leq0}=0$, $\tu_{t\leq0}=0$, $\bw_{t\leq0}=0$.
     \State Set $\tu_\tau=\sum_{i=1}^L M_t^{(i)}\bw_{\tau-i}$, $\tau=0,\dots,t$.
    \For{$\tau=0,1,\dots,t-1$} 
        \State Find an $\alpha$ such that \begin{equation*} \label{eq: PiTraj}
		\begin{bmatrix}
			\tu_{\tau-L+2:\tau+1} \\ \tx_{\tau-L+2}-\bw_{\tau-L+1}
		\end{bmatrix}=\begin{bmatrix}
			H_L(u^d) \\ H_L(x^d)[1,:]
		\end{bmatrix}\alpha
	\end{equation*}
       \State $\tx_{\tau+1}=H_L(x^d)[L,:]\alpha+\bw_{\tau}$
    \EndFor \\
     \Return{$\tx_t$}.
     \end{algorithmic}
\end{algorithm}

\begin{algorithm}
 	\caption{Data-Driven Online Adaptive Control Policy with Clean-Trajectory Input ($\bm{\mA_0}$)}
        \label{alg: noise-free}
 	\begin{algorithmic}[1]
 	\Inputs{$T$, $L$, $N$ such that $2n\leq L\leq N/(m+n+1)$, set $\mathbb{M}$, gradient bound $G$, clean trajectory $\{u^d,x^d\}$.}
 	\State Initialization: $t=0$, $x_0=0$, $x_{t\leq 0}=0$, $u_{-L\leq t\leq T}=0$, $\bw_{t\leq 0}=0$,  $M_{i\leq 0}^{(j)}=0$,$\forall 1\leq j \leq L$, $\lambda=\frac{2LD}{G\sqrt{T}}$, build Hankel matrices $H_L(u^d)$, $H_L(x^d)$.
       \For{$t=0,\dots,T$}
		\State Set $u_t=\sum_{i=1}^L M_t^{(i)}\bw_{t-i}$. 
		\State Receive $x_{t+1}$ and $c_t(u_t,x_t)$. 
		\State Invoke Algorithm \ref{alg: AccNoise}. Calculate $\bw_t$ by \eqref{eq: call AccNoise}. 
		\State Invoke Algorithm \ref{alg: PiTraj}. Simulate $x_t(\bM_t|\{\bw_t\})$ by \eqref{eq: call PiTraj}. 
        \State $f_t(\bM_t)=c_t(u_t,x_t(\bM_t|\{\bw_t\}))$. 
		\State OGD: $\bM_{t+1}=\proj{M}{\bM_t-\lambda\nabla f_t(M_t)}$.
        \EndFor
    \end{algorithmic}
 \end{algorithm}


\section{Data-Driven Online Adaptive Control}
\label{sec: DD-NOC}
In this section, we consider the adversarial online learning problem in a general setting without accessibility to a clean trajectory.
 Compared to the simplified case discussed in Section \ref{sec: previous}, the main challenge of the adversarial online control problem is that we only have access to perturbed data, making it impossible to build a noise-free and non-parametric representation for the system \ref{sys: noise-LTI}. Nevertheless, since Algorithm \ref{alg: noise-free} guarantees a sublinear regret bound, we address this challenge by combining Algorithm \ref{alg: noise-free} with the ETC framework. Specifically, during the exploration stage, the algorithm obtains a sequence $\{u^d,\hx^d\}$ which is sufficiently close to a clean trajectory $\{u^d,x^d\}$ with high probability. As for the commitment stage, the algorithm deploys Algorithm \ref{alg: noise-free} while substituting $\{x^d\}$ for $\{\hx^d\}$. 
Now the main difficulty becomes obtaining a good estimation of a clean trajectory. 

\subsection{Estimation on A Clean Trajectory} \label{subsec: esti}
To obtain a good estimation, we choose $u_t=\{\pm1\}_{i.i.d}^m$ and collect $I_0$ 
independent trajectories starting from $x_0=0$. Define $x^{(i)}=\{x_1^{(i)T},\dots,x_N^{(i)T}\}^T$ as the $i$-th state trajectory, $u^{(i)}=\{u_0^{(i)T},\dots,u_{N-1}^{(i)T}\}^T$ as the control sequence and $\bw^{(i)}=\{\bw_0^{(i)T},\dots,\bw_{N-1}^{(i)T}\}^T$ as the accumulated disturbance respectively. Let $X=\begin{bmatrix}
 x^{(1)} & \cdots & x^{(I_0)} \end{bmatrix}$, $U=\begin{bmatrix} u^{(1)} & \cdots & u^{(I_0)} \end{bmatrix}$ and $\bW=\begin{bmatrix} \bw^{(1)} & \cdots & \bw^{(I_0)} \end{bmatrix}$, then
 \begin{equation}\label{eq: X=}
     X=\Phi U+\bW
 \end{equation}
  where 
 $$\Phi=\begin{bmatrix}
     B & & &\\
     AB & B & &\\
     \vdots & \vdots & \ddots & \\
     A^{N-1}B & A^{N-2}B & \cdots & B 
 \end{bmatrix}.$$

Since $\mbE(U)=0$ and $u_t$ is independent from $w_t$, we right-multiply \eqref{eq: X=} by $U^T$ and take expectation of both side to get
 \begin{equation}
     \mbE(XU^T)=\mbE_U(\Phi UU^T)+\mbE_U(\bW U^T)=I_0\Phi
 \end{equation}
This suggests that $\frac{1}{I_0}XU^T$ is a good estimation of $\Phi$. Given a control sequence $u^d=\{u_0^d,\dots,u_{N-1}^d\}$, $x^d=\Phi u^d$ is a clean trajectory of the system \eqref{sys: noise-LTI}, and thus $\hx^d=\frac{1}{I_0}XU^T u^d$ is close to the $x^d$.
Consequently, we can treat $\{u^d,\hx^d\}$ as an estimation of a clean trajectory, and construct Hankel matrices $H_L(u^d)$ and $H_L(\hx^d)$ as a data-driven representation of the dynamics. 

\subsection{Algorithm Design}
To this end, we are ready to propose the main algorithm for solving adversarial online control problems, see Algorithm \ref{alg: main}. For the inputs of Algorithm \ref{alg: main}, to ensure that Hankel matrices built in Algorithm \ref{alg: main}, Step 12 is full row rank, we require $N-L+1>(m+n)L$ and $L\geq 2n$ \cite[Corollary 2]{2005Anote}. Besides, to avoid redundant discussions on the length of the period, we set $L=\log T$ and $N< T^{2/3}$. The two requirements are captured in the following assumptions.

\begin{assumption}\label{as: full row rank}
    $L=\log T\geq 2n$.
\end{assumption}
\begin{assumption}\label{as: long T}
    $(m+n+1)L\leq N< T^{2/3}$.
\end{assumption}

During the exploration stage (\textit{Stage 1}), the algorithm collects $I_0$ independent trajectories that start from $x_0=0$ with $u_t=\{\pm1\}_{i.i.d.}^m$, and compute $\hx^d$ by $\hx^d=\frac{1}{I_0}XU^T u^d$ with some well-designed $u^d$. To ease the calculation in regret analysis while keeping the condition of persistent excitement of order $L+2n$, we choose $u_t^d\sim\text{Unif}\left(\frac{1}{\sqrt{N}}\mathcal{S}^{m-1}\right)$.

In the commitment stage (\textit{Stage 2}), we follow the same scheme shown in Algorithm \ref{alg: noise-free}. The algorithm is an adaptive ADAC policy, where $u_t$ responds to the estimation of $\bw_t$, denoted as $\hbw_t$, and the parameter $\bM_t$ is updated by OGD. The estimation $\hbw_t$ is calculated by Algorithm \ref{alg: AccNoise} while replacing $H_L(x^d)$ by $H_L(\hx^d)$. To acquire $x_t(\bM_t|\{\hbw_t\})$ which is necessary for OGD, we run Algorithm \ref{alg: PiTraj} while treating $H_L(\hx^d)$ and $\hbw_t$ as the true values. 

\begin{algorithm}
 	\caption{Data-Driven Online Adaptive Control Policy $\bm{\mA}$}
        \label{alg: main}
 	\begin{algorithmic}[1]
 	\Inputs{Time horizon $T$, numbers $I_0$, $L$, $N$, dimension $m$, $n$, set $\mathbb{M}$, gradient bound $G$.}
    \State \textbf{\textit{Stage 1}: Online exploration}
 	\State Initialization: $t=0$, $x_0=0$,
 	\For{$k=0,1,\dots,I_0$}
            \For{$t=0,1,\dots,N$}
               \State Set $u_t^k=\{\pm1\}^{m}_{i.i.d}$ and collect $u_t^k$, $x_{t+1}^k$.
            \EndFor	
        \EndFor
        \State Build matrix $X$ and $U$ as shown in Section \ref{subsec: esti}.
        \State Set $u^d=\{u_0^d,\dots,u_{N-1}^d\}$, $u_t^d\sim\text{Unif}\left(\frac{1}{\sqrt{N}}\mathcal{S}^{m-1}\right)$.
        \State Calculate $\hx^d=\frac{1}{I_0}XU^T u^d$. 
        \State Build Hankel matrices $H(\hx^d)$ and $H(u^d)$. 
	\State \textbf{\textit{Stage 2}: Commitment in Noisy Environment}
        \State Initialization: $t=0$, $T_s=NI_0$, $x_{t\leq 0}=0$, $u_{[-L: T-T_s]}=0$, $\hbw_{t\leq 0}=0$,  $M_{i\leq 0}^{(j)}=0$,$\forall 1\leq j \leq L$, $\lambda=\frac{2LD}{G\sqrt{T}}$.
       \For{$t=0,\dots,T-T_s$}
		\State Set $u_t=\sum_{i=1}^L M_t^{(i)}\hbw_{t-i}$. 
		\State Receive $x_{t+1}$ and $c_t(u_t,x_t)$. 
		\State Calculate $\hbw_t=\text{AccNoise}$$(u_{t-L+2:t}$, $x_{t-L+2}$, $x_{t+1}$, $\bw_{t-L+1}$, $H_L(u^{d})$, $H_L(\hx^d))$.
		\State Calculate $\tx_t(\bM_t)=\text{PiTraj}$$(\hbw_{0:t-1}$,$\bM_t$,$x_{0:L}$, $u_{0:L-1}$, $H_L(u^{d})$, $H_L(\hx^d),t)$. 
        \State $f_t(\bM_t)=c_t(u_t,\tx(\bM_t))$. 
		\State OGD: $\bM_{t+1}=\proj{M}{\bM_t-\lambda\nabla f_t(\bM_t)}$.
        \EndFor
    \end{algorithmic}
 \end{algorithm}

\section{Theoretical Analysis}\label{sec: analysis}
In this section, we analyze the regret bound of the proposed algorithm. The main result is summarized in Theorem \ref{thm: final result}.  
\begin{thm}\label{thm: final result}
     Suppose that all the assumptions hold, then it holds that when $I_0\geq $, for $\forall \delta>0$, the following statement holds with probability at least $1-4\exp(-\delta^2)$,
    $$\verb|rgt|_T(\mA,\Pi^{ADAC})\leq\tmO(C\delta  T^{2/3})$$
    where $C=poly(\rho, \varepsilon, m, n, M, G, N, \|B\|)$. 
\end{thm}

The main ideas of the proof are summarized as follows. First we show that for the system \eqref{sys: noise-LTI}, a non-parametric representation based on the Hankel matrix is equivalent to a linear system model parameterized by some matrices $(H_1,H_2)$, and the true trajectory $\{u_t,x_t\}$, which is a trajectory of $(H_1,H_2)$ subject to $\bw_t$, can also be regarded as a trajectory of some $(\hH_1,\hH_2)$ perturbed by $\hbw_t$. Next, we give the bound of the estimation error of the clean trajectory $\|\hx^d-x^d\|$ and then bound the error $\|\hH_1-H_1\|$, $\|\hH_2-H_2\|$ and $\|\hbw_t-\bw_t\|$ that are caused by $\|\hx^d-x^d\|$. Then we decompose the regret into three parts and bound each separately by using the result of error bounds. Finally, we put the two parts of regret together and obtain the result shown in Theorem \ref{thm: final result}. Most proof details are included in Appendix A-D.

\subsection{$L$-step Representation of LTI Systems}\label{subsec: DD-represent}
We demonstrate the equivalence of system \eqref{sys: noise-LTI}, presented in a 1-step iterative format, to an $L$-step representation. The dynamics of this $L$-step representation can be inferred from $H_L(x^d)$, $H_L(u^d)$ and $\bw_t$.

\begin{lemma}\label{lemma: L-step}
   Let $\{u^d, x^d\}$ be a clean trajectory of \eqref{sys: noise-LTI} and $H_L(u^d)$, $H_L(x^d)$ be associated Hankel matrices. Define $\bu_{t,L-1}=u_{t-L+2:t}$, $H_{ux}=\begin{bmatrix} H_L(u^d) \\ H_L(x^d)[1,:] \end{bmatrix}$ and $H=H_L(x^d)[L,:] H_{ux}^T(H_{ux}H_{ux}^T)^{-1}$. Spilt $H$ into three blocks $H=\left[\begin{array}{c|c|c}
        H_1 & H_0 & H_2
    \end{array} \right]$ with $H_1\in \R^{n\times m(L-1)}$, $H_0\in \R^{n\times m}$ and $H_2\in \R^{n\times n}$, and define $v_t=\bw_t-H_2\bw_{t-L+1}$. Provided that $x_{t\leq0}=0$, $u_{t<0}=0$ and $w_{t<0}=0$, the system \eqref{sys: noise-LTI} is equivalent to the following LTI system
    \begin{equation} \label{sys: Data-LTI}
        x_{t+1}=H_2x_{t-L+2}+H_1\bu_{t,L-1}+v_t
    \end{equation}
\end{lemma}

Thanks to Lemma \ref{lemma: L-step}, we can see that Stage 2 of Algorithm 1 chooses $\bM_t$ to optimize for the cost of the controller on a fictitious linear dynamical system $(\hH_1,\hH_2)$ perturbed by $\hbw_t$, where $(\hH_1,\hH_2)$ follows the same definition of $(H_1,H_2)$ while replacing $x^d$ by $\hx^d$. The following lemma illustrates that the way we generate $\hbw_t$ ensures that the state-control sequence visited by Algorithm \ref{alg: main} coincides with the sequence visited by the regret-minimizing algorithm on the fictitious system.

\begin{lemma}\label{lemma: simulation}
    Suppose that all the assumptions hold. Define $\bu_{t,L-1}=u_{t-L+2:t}$, $\hH_{ux}=\begin{bmatrix} H_L(u^d) \\ H_L(\hx^d)[1,:] \end{bmatrix}$ and $\hH=H_L(\hx^d)[L,:]\cdot \hH_{ux}^T(\hH_{ux}\hH_{ux}^T)^{-1}$. Spilt $\hH$ into three blocks $\hH=\left[\begin{array}{c|c|c}
        \hH_1 & \hH_0 & \hH_2
    \end{array} \right]$ with $\hH_1\in \R^{n\times m(L-1)}$, $\hH_0\in \R^{n\times m}$ and $\hH_2\in \R^{n\times n}$, and define $\hv_t=\hbw_t-\hH_2\hbw_{t-L+1}$. Provided that $x_{t\leq0}=0$, $u_{t<0}=0$ and $w_{t<0}=0$, at any time step $t$ of \textit{Stage 2}, $\{u_t,x_t\}$ satisfies
    \begin{equation} \label{sys: sim-Data-LTI}
        x_{t+1}=\hH_2x_{t-L+2}+\hH_1\bu_{t,L-1}+\hv_t
    \end{equation}
\end{lemma}

\subsection{Estimation Errors and Stability Requirements}\label{subsec: sysid}
We first bound $\delta_d=\|\hx^d-x^d\|$ and then analyze the estimation errors $\|\hH_1-H_1\|,\|\hH_2-H_2\|$ and $\|\hbw_t-\bw_t\|$. 
\begin{lemma}\label{lemmma: delta_d}
     Suppose that all the assumptions hold. Given a control sequence $u^d=\{u_0^d,\dots,u_{N-1}^d\}$ where $u_t^d\sim\text{Unif}(\frac{1}{\sqrt{N}}\mathcal{S}^{n-1})$, let $\hx^d$ and $T_s$ be as Step 11 and 14 of Algorithm \ref{alg: main} separately. Then for any $ \delta>0$, the following statement holds with probability at least $1-4\exp(-\delta^2)$ and some constants $C_1$, $C_2>0$: 
     $$\delta_d\leq\frac{C_x}{\sqrt{T_s}}, ~where $$
    $$C_x=\frac{C_1\varepsilon(\sqrt{mN}+\sqrt{nN}+\delta)+C_2\|B\|N(\sqrt{mN}+\delta)}{(1-\rho)/\sqrt{N}}$$
\end{lemma}
Next we bound the errors $\|\hH_1-H_1\|,\|\hH_2-H_2\|$ and $\|\hbw_t-\bw_t\|$, and discuss the requirements on $T_s$ to ensure stability.
\begin{lemma}\label{lemma: H-hatH}
    Suppose that all the assumptions hold. Define $H_{ux},H,H_1,H_2$ as defined in Lemma \ref{lemma: L-step}, $\hH_{ux},\hH,\hH_1,\hH_2$ as defined in Lemma \ref{lemma: simulation}, and $\delta_H=\|\hH-H\|$. Given a $\delta_d>0$, then as long as $\delta_d\leq\hf$, we have 
    $\delta_H\leq C_H\delta_d$, where 
    $$\begin{cases}
        C_H=(3h-N+1)(8h^3-4h^2+1)-h \\
        h=N+\frac{N\|B\|}{1-\rho}
    \end{cases}
    $$
\end{lemma}
\begin{lemma}\label{lemma: w'-hatw'}
    Suppose that all the assumptions hold. During \textit{Stage 2} of the algorithm, given a $\delta_d>0$, we have, as long as $\delta_d<C_3$,
    $$\begin{cases}
        \|\hbw_t-\bw_t\|\leq C_4\delta_d \\
        \|x_t\|\leq \frac{\|B\|LD}{1-\rho}C_4\delta_d+C_5
    \end{cases}$$
    where 
    $$\begin{cases}
        C_3=\frac{(1-\rho)^2}{2(1-\rho^2)(LD^2+1)+2(1+\rho)\|B\|LD} \\
        C_4=\frac{2\|B\|LD\varepsilon(1+\rho)+2(1-\rho^2)(LD^2+2)\varepsilon}{(1-\rho)^3} \\
        C_5=\frac{(\|B\|LD+1-\rho)\varepsilon}{(1-\rho)^2}
    \end{cases}$$
\end{lemma}
\begin{remark}
It should be noted that since Algorithm \ref{alg: main} produces $u_t$ on the system specialized by the pair $(\hH_1,\hH_2)$, this fictitious system is required to be $(\kappa,\rho_1)$-stable with high probability with some $\kappa>0$ and $0<\rho_1<1$. Without loss of generality, let $\kappa=1$. To meet the stability requirement, $\delta_H$ should satisfy 
\begin{equation}\label{eq: dH+H2<1}
    \delta_H+\|H_2\|< 1
\end{equation}
 Notice that for system \eqref{sys: noise-LTI}, define  $\phi_{L-1}=\begin{bmatrix}
        A^{L-2}B & \cdots & B \end{bmatrix}$, then $x_{t+1}$ can also be represented as 
    \begin{equation}\label{eq: L-step AB}
        x_{t+1}= A^{L-1} x_{t-L+2} + \phi_{L-1}\bu_{t,L-1} + \bw_t-A^{L-1}\bw_{t-L+1}
    \end{equation}
    Since both \eqref{sys: Data-LTI} and \eqref{eq: L-step AB} hold for any clean trajectories, we can get that $H_2=A^{L-1}$ and $\|H_2\|\leq \rho^L$. Therefore, condition \eqref{eq: dH+H2<1} is satisfied as long as $\delta_H\leq\frac{1-\rho^L}{2}$. This condition together with Lemma \ref{lemmma: delta_d}, \ref{lemma: H-hatH} and \ref{lemma: w'-hatw'} implies that in order to establish Lemma \ref{lemma: H-hatH} and \ref{lemma: w'-hatw'}, $T_s$ defined in Algorithm \ref{alg: main}, Step 14 should satisfy
    \begin{equation}\label{eq: Ts satisfies}
        T_s\geq \max\{\frac{4C_H^2C_x^2}{(1-\rho^L)^2},4C_x^2,\frac{C_x^2}{C_3^2}\}.
    \end{equation}
\end{remark}

\subsection{Regret Decomposition and Final Result}\label{subsec: regret decom}
We decompose the regret into $\rgt_1$ in \textit{Stage 1} and $\rgt_2$ in \textit{Stage 2}. We further decompose $\rgt_2$ into $R_1+R_2$ and bound each of them by using the result of error bounds. Finally, we put $\rgt_1$, $R_1$, and $R_2$ altogether to obtain the regret bound.  

Given a $\bM\in\mathbb{M}$ that operates on a linear system specified via $(H_1,H_2)$, define $x_t(\bM|H_1,H_2,\{\bw_t\})$ as the state driven by $u_t(\bM|\{\bw_t\})$ and $c_t(\bM,H_1,H_2,\{\bw_t\})=c_t(u_t(\bM|\{\bw_t\}),x_t(\bM|H_1,H_2,\{\bw_t\}))$, where $u_t(\bM|\{\bw_t\})$ is defined in \eqref{eq: def of M*}.

 In Stage 1 of Algorithm \ref{alg: main}, we do not chase a good policy. Define $T_s=NI_0$, then the regret at this stage is bounded by $\rgt_1\leq \mO(T_s)$.
 
 In Stage 2,  the regret can be decomposed as 
\begin{equation*}\label{eq: rgret<=r1+r2}
    \begin{split}
        &\rgt_2=\sum_{t=0}^{T-T_s-1}c_t(u_t,x_t)-\sum_{t=0}^{T-T_s-1}c_t(\starM,H_1,H_2,\{\bw_t\}) \\
        &\leq \sum_{t=0}^{T-T_s-1} c_t(u_t,x_t)-c_t(\starM,\hH_1,\hH_2,\{\hbw_t\})) \\
        &+ \sum_{t=0}^{T-T_s-1}c_t(\starM,\hH_1,\hH_2,\{\hbw_t\}))-c_t(\starM,H_1,H_2,\{\bw_t\}) \\
        &\defeq R_1+R_2
    \end{split}
\end{equation*}
The following two lemmas ensure that $R_1\leq\tmO(\sqrt{T-T_s})$ and $R_2\leq\tmO(\delta_d(T-T_s))$.
\begin{lemma}\label{lemma: r1}
    Suppose all the assumptions hold. During \textit{Stage 2} of the algorithm, as long as $T_s$ satisfies \eqref{eq: Ts satisfies}, we have $R_1\leq C_6\sqrt{T-T_s}$, where
    $$\begin{cases}
        C_6=2LDG\bigl(\frac{L\varrho}{(1-\varrho)^2}(\frac{\|B\|}{1-\rho}+C_H)(\frac{\varepsilon}{1-\rho}+C_4)\bigr) \\
        \varrho=(\frac{1+\rho^L}{2})^{\frac{1}{L}}
    \end{cases}$$
\end{lemma}
\begin{lemma}\label{lemma: r2}
    Suppose that all the assumptions hold. During \textit{Stage 2} of the algorithm, given a $\delta_d>0$, as long as $T_s$ satisfies \eqref{eq: Ts satisfies}, we have $R_2\leq (C_7+o(1))\delta_d(T-T_s)$, where 
    $$\begin{cases}
        C_7=GC_8+GLDC_4 \\
        C_8= \frac{\|B\|LDC_4}{(1-\rho)^2}+ \frac{LDC_H\varepsilon}{(1-\rho)^2}\bigl(1+\frac{\|B\|}{(1-\varrho)^2}\bigr)+C_4 \\
        \varrho=(\frac{1+\rho^L}{2})^{\frac{1}{L}}
    \end{cases}$$
\end{lemma}

Note that if $\{\hx^d\}$ in Algorithm\ref{alg: main} is exactly a clean state trajectory, then $\delta_d=0$ and $T_s=0$. By Lemma \ref{lemma: r1} and Lemma \ref{lemma: r2} we know that $\rgt_T(\mA,\Pi^{ADAC})\leq 0+R_1+0\leq\tmO(\sqrt{T})$, which is the result of Theorem \ref{thm: sqrt-T regret}. In typical cases, put Lemma \ref{lemma: r1}, Lemma \ref{lemma: r2} and Lemma \ref{lemmma: delta_d} altogether and we obtain
\begin{subequations}\label{eq: rgt1+R1+R2}
    \begin{align}
        &\texttt{rgt}_T(\mA,\Pi^{ADAC}) \\
        =&\rgt_1+R_1+R_2 \\
        \leq &\tmO(T_s+C_6\sqrt{T-T_s}+C_7C_x\frac{T-T_s}{\sqrt{T_s}})
    \end{align}
\end{subequations}

By \cite[Chapter 10]{hazan2022introduction}, \eqref{eq: rgt1+R1+R2}  yields the result of Theorem \ref{thm: final result} by setting $T_s=NI_0=[\max \{T^{2/3},C_9\}]$, where $C_9$ is the right-hand side of \eqref{eq: Ts satisfies}.
    

\section{Data-driven control with output feedback} \label{sec: output}
An advantage of our algorithm is that it can be directly applied to the output feedback case with minimal modifications. We first formulate the problem, then show how Algorithm \ref{alg: main} works with output sequence, and finally state the regret bound, which remains to be $\tmO(T^{2/3})$.

\subsection{Problem Formulation}
Consider the following adversarial online control problem with output feedback
\begin{subequations}\label{sys: LTI output}
    \begin{align}
        \min J&=\sum_{t=0}^{T-1} c_(u_t,y_t) \\
        \text{s.t.}\quad &x_{t+1}=Ax_t+Bu_t+w_t \\
        &y_t=Cx_t+e_t,
    \end{align}
\end{subequations}
where $x_t\in\R^n$ is the state, $u_t\in\R^m$ is the control, $w_t\in\R^n$ is the disturbance on the state,  $y_t\in \R^{p}$ is the output, and $e_t\in\R^p$ is the measurement disturbance. In this case, both $w_t$ and $e_t$ satisfy Assumption \ref{as: bounded_w}, $(A,B)$ satisfies Assumption \ref{as: AB}, $c_t$ satisfies Assumption \ref{as: cost} and $C\in \R^{p\times n}$ is bounded to keep the $(\kappa,\rho)$ stability of the system.

At each time step $t$, the learner plays $u_t$ and then observes $y_{t+1}$ and the cost $c_t(u_t,y_t)$. The system matrices $(A,B,C)$ are unknown. In the procedure of model-based algorithm \cite{simchowitz2020improper}, since $(A,B,C)$ are not detectable with output feedback, it requires identifying a sequence of Markov operator $\hat{G}^{\{1:h\}}$, which is complicated and time-consuming. On the contrary, Algorithm \ref{alg: main} can be directly extended to this problem while bypassing the identification of any parameters. 

\subsection{Application of Algorithm \ref{alg: main}}
To begin with, for clean trajectories $\{u^d,y^d\}$ and $\{\hat{u},\hat{y}\}$, the following statement holds as the corollary of the fundamental lemma.

\begin{corollary}[Corollary of Fundamental Lemma]\label{coro: Hankel_y}
Suppose $\{u^d,y^d\}$ is a noise-free trajectory of the system (\ref{sys: LTI output}), where $\{u^d\}$ is persistently exciting of order $L+2n$. Then, $\{\hat{u},\hat{y}\}=\{(\hat{u}_1,\hat{y}_1),\dots,(\hat{u}_L,\hat{y}_L)\}$ is a noise-free trajectory of system (\ref{sys: noise-LTI}) if and only if there exists $\alpha\in\mathbb{R}^{N-L+1}$ such that
\begin{equation} \label{eq: DD-Represent_y}
    \begin{bmatrix}
        H_L(u^d) \\ H_L(y^d)
    \end{bmatrix}\alpha=\begin{bmatrix}
        \hat{u} \\ \hat{y}
    \end{bmatrix}
\end{equation}
\end{corollary}

This corollary implies that the precise non-parametric representation of the system \eqref{sys: LTI output} can be built upon Hankel matrices of a clean output trajectory $\{H_L(u^d),H_L(y^d)\}$. 

Now define $\bw_t^{o}=e_t+\sum_{i=0}^{t}{CA^iw_{t-i}}$, which can be treated as the \textit{observation} of the accumulated disturbance $\bw_t$, then $y_t$ can be expressed as
\begin{equation}\label{eq: y_t=}
    y_t=\sum_{i=1}^t{CA^{i-1}Bu_{t-i}}+\bm{w}_{t-1}^{o}
\end{equation}
The first item of on the RHS of \eqref{eq: y_t=} can be treated as the output of a clean trajectory driven by $u_t$. By Corollary \ref{coro: Hankel_y}, there exists an $\alpha\in\mathbb{R}^{N-L+1}$ such that
\begin{equation} \label{eq: u';y'=H;H}
    \begin{bmatrix}
        u_{t-L+1:t} \\ y_{t-L+1:t}-\accw_{t-L:t-1}
    \end{bmatrix}=\begin{bmatrix}
        H_L(u^d) \\ H_L(y^d)
    \end{bmatrix}\alpha
\end{equation}

Therefore, we can calculate $\bw_t^{o}$ by Algorithm \ref{alg: AccNoise} while replacing $H_x$ by $H_y=H_L(y^d)$ and $x_{t-L+2}, x_{t+1}$ by $y_{t-L+2}, y_{t+1}$, provided that $\{y^d\}$ is clean. To this end, we modify ADAC as an observed disturbance-action controller (ODAC) as
$$u_t^{(ODAC)}=\sum_{i=1}^L M^{(i)}\bw^{o}_{t-i},~~M\in\mathbb{M},$$
and choose ODAC as a reference policy class. \cite[Chapter 8,9]{hazan2022introduction} provides further discussion on the rationality of the ODAC. The regret in this problem is therefore specified as
\begin{equation}
\begin{split}
     &\rgt_T(\mA_y,\Pi)=\sum_{t=0}^{T-1}c_t(u_t,y_t) \\
     &-\sum_{t=0}^{T-1}c_t(u_t(\starM|\{\bw_t^{o}\}),y_t(\starM|\{\bw_t^{o}\}))
\end{split}
\end{equation}
where 
$$\begin{cases}
    \starM=\arg\min_{\bM\in\mathbb{M}}\sum_{t=0}^{T-1}{c(u_t(\bM|\{\bw_t^{o}\}),y_t(\bM|\{\bw_t^{o}\}))} \\
    y_t(\bM|\{\bw_t^{o}\})=Cx_t(\bM|\{\bw_t^{o}\}) \\
    x_{t+1}(\bM|\{\bw_t^{o}\})=Ax_{t}(\bM|\{\bw_t^{o}\})+Bu_t(\bM|\{\bw_t^{o}\})+w_t \\
    u_t(\bM|\{\bw_t^{o}\})=\sum_{i=1}^L M^{*(i)}\bw_{t-i}^{o}
\end{cases}$$

In the case where we only have noise-corrupted output data, we can estimate a trajectory $\{\hat{y}^d\}$ by the same strategy as shown in \textit{Stage 1}, Algorithm \ref{alg: main}. Here we still choose $u_t=\{\pm1\}_{i.i.d}^m$ and collect $I_0$ independent trajectories starting from $x_0=0$. Define $y^{(i)}=\{y_1^{(i)T},\dots,y_N^{(i)T}\}^T$ as the $i$-th state trajectory, $u^{(i)}=\{u_0^{(i)T},\dots,u_{N-1}^{(i)T}\}^T$ as the control sequence and $\bw^{o(i)}=\{\bw_0^{o(i)T},\dots,\bw_{N-1}^{o(i)T}\}^T$ as the disturbance sequence respectively. Let $Y=\begin{bmatrix}
 y^{(1)} & \cdots & y^{(I_0)} \end{bmatrix}$, $U=\begin{bmatrix} u^{(1)} & \cdots & u^{(I_0)} \end{bmatrix}$ and $\bW^{o}=\begin{bmatrix} \bw^{o(1)} & \cdots & \bw^{o(I_0)} \end{bmatrix}$ then we have
 \begin{equation}\label{eq: Y=}
     Y=\Phi_y U+\bW^{o}
 \end{equation}
  where 
 $$\Phi_y=\begin{bmatrix}
     CB & & &\\
     CAB & CB & &\\
     \vdots & \vdots & \ddots & \\
     CA^{N-1}B & CA^{N-2}B & \cdots & CB 
 \end{bmatrix}$$
By right-multiplying $U^T$ to both sides of \eqref{eq: Y=} and taking expectation, we can find that $\hPhi_y=\frac{1}{I_0}YU^T$ is a good estimation of $\Phi_y$. An approximated clean trajectory can be constructed by choosing $u^d=\{u_0^d,\dots,u_{N-1}^d\}$, $u_t^d\sim\text{Unif}\left(\frac{1}{\sqrt{N}}\mathcal{S}^{m-1}\right)$ and setting $\hy^d=\hPhi y^d$. By following the same steps of the proof of Lemma \ref{lemmma: delta_d}, we can show the estimation error $\|\hy^d-y^d\|$ is on the same level as $\|\hx^d-x^d\|$ shown in Lemma \ref{lemmma: delta_d}.  

With $\{u^d,\hy^d\}$ in hand, we treat $\{u^d,\hy^d\}$ as a clean trajectory and operate an adaptive ODAC, just as \textit{Stage 2}, Algorithm \ref{alg: main} does. The controller is with respect to the estimation of $\bw_t^{o}$, denoted as $\hbw_t^{o}$, which is calculated by Algorithm \ref{alg: AccNoise} while replacing $H_L(x^d)$ by $H_L(\hy^d)$. The parameter $\bM_t$ is updated by OGD. In order to acquire $y_t(\bM_t|\{\hbw_t^{o}\})$ which is necessary for OGD, we run Algorithm \ref{alg: PiTraj} by treating $H_L(\hy^d)$ and $\hbw_t^{o}$ as the true values. 

We are ready to provide the output-feedback edition of Algorithm \ref{alg: main}, denoted as Algorithm \ref{alg: yt}. Since the two algorithms are almost the same, only changing state variables $x$ to output variables $y$ and $\hbw_t$ to $\hbw_t^{o}$, we include Algorithm \ref{alg: yt} in Appendix \ref{subsec: alg yt}.
Following the same steps as in Section \ref{sec: analysis}, we establish an $\tmO(T^{2/3})$ regret bound for Algorithm \ref{alg: yt}.
\begin{thm}\label{thm: output result}
     Suppose that all the assumptions hold, then it holds that when $I_0\geq $, for $\forall \delta>0$, the following statement holds with probability at least $1-4\exp(-\delta^2)$,
    $$\texttt{rgt}_T(\mA_y,\Pi^{ODAC})\leq\tmO(C_y\delta\cdot T^{2/3})$$
    where $C_y=poly(\rho, \varrho, m, p, M, G, N, \|B\|,\|C\|)$.
\end{thm}

\begin{remark}
Theorem \ref{thm: output result} tells that the regret guarantee of our algorithm is at the same level as \cite{simchowitz2020improper}, which solves the adversarial online control problem with output feedback by a model-based algorithm. However, the computational cost of our algorithm is lower. The most time-consuming calculation in Algorithm \ref{alg: yt} is solving a linear equation in the form of $$\begin{bmatrix}
			\bu \\ y_{t-L+2}-\hbw^o_{t-L+1}
		\end{bmatrix}=\begin{bmatrix}
			H_L(u^d) \\ H_L(y^d)[1,:]
		\end{bmatrix}\alpha,$$ which can be done efficiently. As for the model-based algorithm, the heaviest computation task is solving an optimization problem formulated as
$$\{G_1,\dots,G_N\}=\arg\min \sum_{t=N+1}^{I_0}\|y_t-\sum_{i=1}^N G_iu_{t-i}\|^2,$$
where $G\in \R^{p\times m}$ is a matrix. The computational cost of solving it is $\tilde{\Omega}(I_0Npm^3)$, depending on the optimization method. When $L$ is relatively small, our algorithm has a lower computational cost.  
\end{remark}

\section{Conclusions}\label{sec: conclu}
In this paper we propose an efficient data-driven algorithm for control an unknown linear dynamical system in the face of adversarial disturbances and adversarial convex loss functions. The data-driven representation is based on Hankel matrices in behavioral systems theory and the controller is chosen as an adaptive ADAC whose parameter is updated by OGD. Our algorithm coordinates these parts under the framework of ETC, and guarantees an $\tmO(T^{2/3})$ regret bound with high probability. Moreover, our algorithm is versatile, extending naturally to scenarios where only output feedback is available. A limitation of this work is that it needs to collect multiple trajectories during the exploration stage. Whether a single perturbed trajectory is enough for this data-driven scheme is considered as a future topic. 


\bibliographystyle{unsrt}        
\bibliography{paper}          

\appendix
\appendix
\label{sec: appendix}

\section{Technical Lemmas}
This section states three classical conclusions in high-dimensional probability and linear algebra that are used in the proof.

\begin{lemma}\label{lemma: random matrix}
    \text{\cite[Chapter 4]{vershynin2018high}} Let $A=\{a_{ij}\}_{m\times n}\in\R^{m\times n}$ where each element $a_{ij}$ is independently drown from a sub-Gaussian distribution $G_{ij}$ with variance proxy $\sigma^2_{ij}$, then it holds that for $K=\max_{i,j} |\sigma_{ij}|$, some constant $C>0$ and any $\delta>0$,
    $$\mathbb{P}\left(\|A\|\leq CK(\sqrt{m}+\sqrt{n}+\delta)\right)\geq 1-2\exp(-\delta^2).$$
\end{lemma}

\begin{lemma}\label{lemma: HuxFullrank}
    \cite[Corollary 2]{2005Anote} Suppose that $\{u^d,x^d\}$ is a clean trajectory of \eqref{sys: noise-LTI} and $\{u^d\}$ is persistently exciting of order $L+2n$, then $H_{ux}$ defined by Lemma \ref{lemma: L-step} is full row rank.
\end{lemma}

\begin{lemma}\label{lemma: inv A+dA}
    \cite{miller1981on} Given an invertible matrix $A\in\R^{n\times n}$, let $\sigma_i$, $i=1,2,\dots,n$ be its eigenvalues. Then for any $\delta_A\in\R^{n\times n}$ such that $\delta_A\neq -A$, it holds that,
    \begin{equation*}
        (A+\delta_A)^{-1}=A^{-1}-A^{-1}\delta_A A^{-1}(I+\delta_AA^{-1})^{-1}
    \end{equation*}
\end{lemma}

\section{Proof and Analysis for Section \ref{subsec: DD-represent}}
\subsection{Proof of Lemma \ref{lemma: L-step}}
By Algorithm 1, 
\begin{equation}\label{eq: FindAccNoise}
    \bw_{t}= x_{t+1}-H_L(x^d)[L,:]\alpha,
\end{equation}
where $\alpha$ satisfies
\begin{equation}\label{eq: FindAlpha}
    \begin{bmatrix}
			u_{t-L+2:t+1} \\ x_{t-L+2}-\bw_{t-L+1}
		\end{bmatrix}=H_{ux}\alpha
\end{equation}

Although typically $\alpha$ is not unique, it's easy to find the one with the smallest norm when $H_{ux}$ is full row rank as 
\begin{equation}\label{eq: psudo inv alpha}
    \alpha=H_{ux}^T(H_{ux}H_{ux}^T)^{-1}\begin{bmatrix}
			u_{t-L+2:t+1} \\ x_{t-L+2}-\bw_{t-L+1}
		\end{bmatrix}
\end{equation}
Plug \eqref{eq: psudo inv alpha} into \eqref{eq: FindAccNoise} and apply the notation $H=\left[\begin{array}{c|c|c} H_1 & H_0 & H_2 \end{array} \right]$, we obtain
\begin{equation}\label{eq: xt+1=H1H0H2}
    x_{t+1}=H_1u_{t-L+2:t}+H_0u_{t+1}+H_2(x_{t-L+2}-\bw_{t-L+1})+\bw_t
\end{equation}

It should be noted that whatever the value of $u_{t+1}$ is in \eqref{eq: FindAlpha}, it does not affect states in this trajectory. Therefore, we set $u_{t+1}=0$ and plug it in \eqref{eq: xt+1=H1H0H2} to get 
\begin{equation}
    x_{t+1}=H_1u_{t-L+2:t}+H_2x_{t-L+2}+(\bw_t-H_2\bw_{t-L+1})
\end{equation}
which is exactly the conclusion of Lemma \ref{lemma: L-step}.

\subsection{Proof of Lemma \ref{lemma: simulation}}
    Since \textit{Stage 2} invokes Algorithm \ref{alg: AccNoise} to calculate $\hbw_t$, by Lemma \ref{lemma: L-step} we know that 
    \begin{equation} \label{eq: get hbw}
        \hbw_t= x_{t+1}-\hH_2x_{t-L+2}-\hH_1\bu_{t,L-1}+\hH_2\hbw_{t-L+1}
    \end{equation}
    Move $x_{t+1}$ to the LHS of \eqref{eq: get hbw} and the other parts to the RHS, then it becomes \eqref{sys: sim-Data-LTI} in Lemma \ref{lemma: simulation}.

\section{Proof and Analysis for Section \ref{subsec: sysid}}
\subsection{Proof of Lemma \ref{lemmma: delta_d}}
To begin with, we derive the concentration inequality of $\bW U^T$. Define $\varpi_{ij}$ as elements of $\bW U^T$, since $u_t$ is independent from each other, it's enough to derive $\|\varpi_{11}\|_{\psi_2}$ and generalize to any $\varpi_{ij}$. Calculate $\bW U^T$ and we have
    \begin{equation}
        \varpi_{11}=\sum_{i=0}^{I_0} \bw_{01}^{(i)}u_{01}^{(i)},\quad u_{01}^{(i)}=\begin{cases}
            1\quad p=\frac{1}{2}; \\
            -1\quad p= \frac{1}{2} 
        \end{cases}
    \end{equation}
    Define $\tilde{\bw}_{01}=\begin{bmatrix} \bw_{01}^{(1)} & \cdots & \bw_{01}^{(I_0)} \end{bmatrix}$ and $\tilde{u}_{01}=\begin{bmatrix} u_{01}^{(1)} & \cdots & u_{01}^{(I_0)} \end{bmatrix}$. By the definition of $\bw_t$, we have that 
    \begin{equation}
        |\bw_{01}^{(i)}|\leq \|\bw_0^{(i)}\|\leq \varepsilon\sum_{k} \rho^k \leq \frac{\varepsilon}{1-\rho}
    \end{equation}
    \begin{equation}
        \Rightarrow\|\tilde{\bw}_{01}\|^2\leq \frac{I_0\varepsilon^2}{(1-\rho)^2}
    \end{equation}
    Using Hoeffding's Inequality, we have for $\forall\delta>0$
    \begin{subequations}\label{eq: conc of WU11}
        \begin{align}
            &\mathbb{P}\left(\bigl|\varpi_{11}\bigr|\geq\delta\right) \leq 2\exp\bigl(-\frac{\delta^2}{2\|\tilde{\bw}_{01} \|^2}\bigr) \\
            \leq&  2\exp\bigl(-\frac{\delta^2(1-\rho)^2}{2I_0\varepsilon^2}\bigr)
        \end{align}        
    \end{subequations}
This means that $\|\varpi_{ij}\|_{\psi_2}=\frac{C\varepsilon\sqrt{I_0}}{1-\rho}$, $\forall i\in\{1,\dots,nN\}$, $j\in\{1,\dots,mN\}$ with some constant $C$. Therefore,
\begin{equation}\label{eq: WUpsi2}
    \|\frac{1}{I_0}\varpi_{ij}\|_{\psi_2}=\frac{C\varepsilon}{(1-\rho)\sqrt{I_0}}
\end{equation}
Plug \eqref{eq: WUpsi2} into Lemma \ref{lemma: random matrix} and we have the following statement holds with probability at least $1-2\exp(-\delta^2)$ and some constant $C_1>0$,
\begin{equation}\label{eq: WU^T}
    \|\frac{1}{I_0}\bW U^T\|\leq \frac{C_1\varepsilon(\sqrt{mN}+\sqrt{nN}+\delta)}{(1-\rho)\sqrt{I_0}}
\end{equation}

Next, we move to the concentration inequality of $UU^T$. Define $\nu_{ij}$ as elements of $UU^T$. Notice that $\bigl(u_{tk}^{(i)}\bigr)^2=1$ for any $t\leq N$, $k\leq m$ and $i\leq I_0$, so we have $\nu_{ii}=I_0$ for any $i\in\{1,\dots,mN\}$. For the elements not on the diagonal, since $u_t$ is independent of each other, it's enough to derive $\|\nu_{12}\|_{\psi_2}$ and generalize to any $\nu_{ij}$. Calculating $UU^T$ points to
    \begin{equation}
        \nu_{12}=\sum_{i=1}^{I_0} u_{01}^{(i)}u_{02}^{(i)},\quad u_{01}^{(i)}u_{02}^{(i)}=\begin{cases}
            1\quad p=\frac{1}{2}; \\
            -1\quad p= \frac{1}{2} 
        \end{cases},
    \end{equation}
which implies that each $u_{01}^{(i)}u_{02}^{(i)}$ is a symmetric Bernoulli random variable. By Hoeffding's Inequality, we obtain
    \begin{equation}
        \mathbb{P}{\bigl|\nu_{ij}\bigr|\geq\delta}\leq 2\exp\bigl(-\frac{\delta^2}{2I_0}\bigr)
    \end{equation}
which indicates that $\pto{\nu_{ij}}=C\sqrt{I_0}$ with some constant $C$, $\forall i,j\in\{1,\dots,mN\}$, $i\neq j$. Let $\frac{1}{I_0}UU^T=I+V$, then we have 
\begin{equation}\label{eq: UUpsi2}
    \pto{v_{ij}}=\begin{cases}
        0\quad i=j; \\
        \frac{C}{\sqrt{I_0}}\quad i\neq j
    \end{cases}
\end{equation}

Plugging \eqref{eq: UUpsi2} into Lemma \ref{lemma: random matrix} yields that, for any $\delta>0$ and some constant $C_2$, with probability of at least $1-2\exp(-\delta^2)$,
\begin{equation}\label{eq: UUnorm}
    \|\frac{1}{I_0}UU^T-I\|=\|V\|\leq \frac{C_2(\sqrt{mN}+\delta)}{\sqrt{I_0}},
\end{equation}

Define $\hPhi=\frac{1}{I_0}XU^T$. Now we consider combining \eqref{eq: UUnorm} and \eqref{eq: WU^T} to bound $\|\hat{\Phi}-\Phi\|$ as well as $\delta_d$, where $\Phi$ is defined in \eqref{eq: X=}. Multiplying the left-hand side of \eqref{eq: UUnorm} by $\Phi$ we obtain
\begin{equation}
    \|\Phi\cdot\frac{1}{I_0}UU^T-\Phi\|\leq \|\Phi\|\cdot \|\frac{1}{I_0}UU^T-I\|
\end{equation}

Therefore, we have for any $\delta>0$
\begin{equation}\label{eq: Phinorm}
    \begin{split}
        &\mathbb{P}\left(\|\Phi\cdot\frac{1}{I_0}UU^T-\Phi\|\geq \|\Phi\|\frac{C(\sqrt{mN}+\delta)}{\sqrt{I_0}}\right) \\
        \leq &\mathbb{P}\left(\|\Phi\|\cdot \|\frac{1}{I_0}UU^T-I\|\geq \|\Phi\|\frac{C(\sqrt{mN}+\delta)}{\sqrt{I_0}}\right)\\
        =&\mathbb{P}\left(\|\frac{1}{I_0}UU^T-I\|\geq \frac{C(\sqrt{mN}+\delta)}{\sqrt{I_0}}\right) \\
        \leq &2\exp(-\delta^2)
    \end{split}
\end{equation}

Consider events $E_1= \{\|\frac{1}{I_0}\bW U^T\|\leq \frac{C_1\varepsilon(\sqrt{mN}+\sqrt{nN}+\delta)}{(1-\rho)\sqrt{I_0}}\}$ and $E_2=\{\|\Phi\cdot\frac{1}{I_0}UU^T-\Phi\|\leq \|\Phi\|\frac{C_2(\sqrt{mN}+\delta)}{\sqrt{I_0}}\}$ with some $C1, C_2>0$. By \eqref{eq: Phinorm} and \eqref{eq: WU^T} we have 
\begin{equation}\label{eq: E1capE2}
    \mathbb{P}\left({E_1\cap E_2}\right)\geq 1-4\exp(-\delta^2)
\end{equation}

Also, note that 
\begin{subequations}\label{eq: decom phi-phi^}
    \begin{align}
        &\|\hat{\Phi}-\Phi\|=\|\Phi\cdot\frac{1}{I_0}UU^T-\Phi+\bW U^T\| \\
        \leq &\|\Phi\cdot\frac{1}{I_0}UU^T-\Phi\|+ \|\bW U^T\|
    \end{align}
\end{subequations}

Combining \eqref{eq: E1capE2} and \eqref{eq: decom phi-phi^}, we have that with probability at least $1-4\exp(-\delta^2)$
\begin{equation}\label{eq: phi-phi^}
    \begin{split}
        \|\hat{\Phi}-\Phi\|\leq &\frac{C_1\varepsilon(\sqrt{mN}+\sqrt{nN}+\delta)}{(1-\rho)\sqrt{I_0}} \\
        &+\frac{C_2\|B\|N(\sqrt{mN}+\delta)}{(1-\rho)\sqrt{I_0}}
    \end{split}
\end{equation}

By the definition of $\hx^d$ we know that with probability at least $1-4\exp(-\delta^2)$ we have
\begin{equation}\label{eq: x^d-hatx^d}
\|\hat{x}^d-x^d\|\leq \|u^d\|\|\hPhi-\Phi\|
\end{equation}
Combining \eqref{eq: phi-phi^} and \eqref{eq: x^d-hatx^d}, in view of $T_s=(N+1)I_0$ and $\|u^d\|=\sqrt{\sum_{t=0}^{N-1}(u_t^d)^2}=1$ when $u_t^d\sim\text{Unif}\left(\frac{1}{\sqrt{N}}\mathcal{S}^{m-1}\right)$, we arrive at the conclusion of Lemma \ref{lemmma: delta_d}.  

\subsection{Proof of Lemma \ref{lemma: H-hatH}}
Recall that $H=H_L(x^d)[L,:]\cdot H_{ux}^T(H_{ux}H_{ux}^T)^{-1}$ and $\hH=H_L(\hx^d)[L,:]\cdot \hH_{ux}^T(\hH_{ux}\hH_{ux}^T)^{-1}$. Define $\delta_1=H_L(\hx^d)[L,:]-H_L(x^d)[L,:]$, $\delta_2=\hH_{ux}^T-H_{ux}^T$ and $\delta_3=(\hH_{ux}\hH_{ux}^T)^{-1}-(H_{ux}H_{ux}^T)^{-1}$. Next, we derive the bound on $\delta_1$, $\delta_2$ and $\delta_3$.

\begin{itemize}
    \item Bound of $\|\delta_1\|$. Notice that $H_L(\hx^d)[L,:]$ is part of $\hx^d$, so as $H_L(\hx^d)[L,:]$, we have 
    \begin{equation} \label{eq: bound delta_1}
        \|\delta_1\|\leq \|x^d-\hx^d\|=\delta_d
    \end{equation}
\end{itemize}
\begin{itemize}
    \item Bound of $\|\delta_2\|$. Notice that 
    \begin{equation*}
    \begin{split}
        \delta_2&=\begin{bmatrix} H_L(u^d) \\ H_L(\hx^d)[1,:] \end{bmatrix}^T-\begin{bmatrix} H_L(u^d) \\ H_L(x^d)[1,:] \end{bmatrix}^T \\
        &=\begin{bmatrix} 0 \\ H_L(\hx^d)[1,:]- H_L(x^d)[1,:]\end{bmatrix}^T
    \end{split}
    \end{equation*}
    and $H_L(\hx^d)[1,:]$ is part of $\hx^d$, so as $H_L(\hx^d)[1,:]$. Consequently,
    \begin{equation} \label{eq: bound delta_2}
        \|\delta_2\|\leq \|x^d-\hx^d\|=\delta_d
    \end{equation}
\end{itemize}
\begin{itemize}
    \item Bound of $\|\delta_3\|$. Define $P=H_{ux}H_{ux}^T$, $\hP=\hH_{ux}\hH_{ux}^T$. First we bound $\delta_P=\|P-\hP\|$ as
    \begin{equation*}
        \begin{split}
            \delta_P=&\|(H_{ux}+\delta_2)(H_{ux}+\delta_2)^T-H_{ux}H_{ux}^T\| \\
            =&\|H_{ux}\delta_2^T+\delta_2H_{ux}^T+\delta_2\delta_2^T\| \\
            \leq& 2\|H_{ux}\|\delta_d+\delta_d^2 \quad\quad (\text{Apply \eqref{eq: bound delta_2}})
        \end{split}
    \end{equation*}
    Then applying Lemma \ref{lemma: inv A+dA} we get
    \begin{subequations}
        \begin{align}
            &\|\delta_3\|=\|(P+\delta_P)^{-1}-P^{-1}\| \\
            =&\|P^{-1}\delta_PP^{-1}(I+\delta_PP^{-1})^{-1}\| \\
            \leq&\frac{\|\delta_P\|}{\|I+\delta_PP^{-1}\|\cdot\|P\|^2} \\ \leq&\frac{2\|H_{ux}\|\delta_d+\delta_d^2}{(1-\frac{2\|H_{ux}\|\delta_d+\delta_d^2 }{\|H_{ux}\|^2})\|H_{ux}\|^4}
        \end{align}
    \end{subequations}
    On the one hand
    \begin{equation*}
        \begin{split}
            \|H_{ux}\|\geq \|H_L(u^d)\|\geq \sqrt{\frac{L(N-L+1)}{N}}\geq 1
        \end{split}
    \end{equation*}
    On the other hand
    \begin{equation*}
        \begin{split}
            \|H_{ux}\|\leq N+\|\Phi\|
            \leq N+\|B\|\sum_{i=0}^{N-1}(N-i)\rho^i
        \end{split}
    \end{equation*}
    \begin{equation}\label{eq: bound Hux}
        \Rightarrow 1\leq\|H_{ux}\|\leq N+\frac{N\|B\|}{1-\rho}
    \end{equation}
\end{itemize}
By \eqref{eq: bound Hux} as well as $\delta_d\leq\hf$, we have
\begin{subequations}
        \begin{align}
            &\|\delta_3\|\leq\frac{2\|H_{ux}\|\delta_d+\delta_d^2}{(1-\frac{2\|H_{ux}\|\delta_d+\delta_d^2 }{\|H_{ux}\|^2})\|H_{ux}\|^4} \\
             \leq &4N^2(1+\frac{\|B\|}{1-\rho})^2(2N(1+\frac{\|B\|}{1-\rho})+1)\delta_d
        \end{align}
    \end{subequations}

Define $h=N(1+\frac{\|B\|}{1-\rho})$. Put $\delta_1$, $\delta_2$ and $\delta_3$ together and apply \eqref{eq: bound Hux} as well as the condition that $\delta_d\leq\hf<1$, then we have 
\begin{subequations}
    \begin{align*}
        &\|\hH-H\| \\
        =&\|(H_L(x^d)[L,:]+\delta_1)(H_{ux}^T+\delta_2)((H_{ux}H_{ux}^T)^{-1}+\delta_3) \\
        &-H_L(x^d)[L,:]\cdot H_{ux}^T(H_{ux}H_{ux}^T)^{-1}\| \\
        \leq&  \biggl(\frac{\|x^d\|+h+1}{\|H_{ux}\|^2}+(h+1)(8h^3+4h^2) \\
        &+(\|x^d\|+h)(8h^3+4h^2)\biggr)\delta_d\\
        \leq& \left((3h-N+1)(8h^3-4h^2+1)-h\right)\delta_d
    \end{align*}
\end{subequations}
According to the definition, we have $$\hH-H=\left[\begin{array}{c|c|c}
        \hH_1-H_1 & \hH_0-H_0 & \hH_2-H_2
    \end{array} \right].$$  
   It follows that $$\|\hH_1-H_1\|,\|\hH_2-H_2\|\leq \|\hH-H\|$$
    which completes the proof of Lemma \ref{lemma: H-hatH}.

\subsection{Proof of Lemma \ref{lemma: w'-hatw'}}
We use mathematical induction to prove this lemma. Suppose that there exist $\beta,\gamma\in\R$ such that $\|\hbw_{\tau-1}-\bw_{\tau-1}\|\leq\beta$ and $\|x_\tau\|\leq\gamma$ hold for all $\tau\leq t$. When $t=0$, the initialization tells us
\begin{equation}\label{eq: |wt| initial}
    \begin{cases}
        \|\hbw_{t<0}-\bw_{t<0}\|=0\leq\beta \\
        \|x_{t\leq0}\|=0\leq\gamma
    \end{cases}
\end{equation}

 When $t\to t+1$: 
\begin{itemize}
    \item Since $x_{t+1}=Ax_t+Bu_t+w_t$ and $u_t=\sum_{i=1}^L M_t^{(i)}\hbw_{t-i}$, we have for any $1\leq i\leq L$,
    \begin{equation*} \label{eq: hatw'<=}
       \|\hbw_{t-i}\|\leq \|\bw_{t-i}\|+\|\hbw_{t-i}-\bw_{t-i}\|\leq \frac{\varepsilon}{1-\rho}+\beta 
    \end{equation*}
    and thus
    \begin{subequations}\label{eq: x_t+1}
        \begin{align}
            \|x_{t+1}\|=\|Ax_t+B\sum_{i=1}^L M_t^{(i)}\hbw_{t-i}+w_t\| \\
            \leq \rho\gamma+\|B\|LD(\beta+\frac{\varepsilon}{1-\rho})+\varepsilon\leq\gamma
        \end{align}
    \end{subequations}
\end{itemize}

\begin{itemize}
    \item To calculate $\hbw_t$, recall that $\hbw_t=x_{t+1}-H_L(\hx^d)[L,:]\alpha$, where $\alpha$ satisfies \eqref{eq: psudo inv alpha}. Define 
    $$\begin{cases}
        \tbw_k^{(t)}=x_{t-L+k+1}-H_L(\hx^d)[k,:]\alpha, ~1\leq k\leq L \\
        \bbw_k^{(t)}=x_{t-L+k+1}-A(x_{t-L+k}-\tbw_{k-1}^{(t)})-Bu_{t-L+k},
    \end{cases}$$
    then $\tbw_1^{(t)}=\hbw_{t-L+1}$ and $\tbw_L^{(t)}=\hbw_t$. When $k\geq2$,
    \begin{equation}\label{eq: tbw-bbw}
        \begin{split}
            &\|\tbw_k^{(t)}-\bbw_k^{t}\| \\
            \leq &\|AH_L(\hx^d)[k-1,:]+BH_L(u^d)[k-1,:]-H_L(\hx^d)[k,:]\|\cdot \|\alpha\| \\
            \leq & \|AH_L(\hx^d)[k-1,:]+BH_L(u^d)[k-1,:]-H_L(x^d)[k,:]\|\cdot \|\alpha\| \\
            &+ \|H_L(x^d)[k,:]-H_L(\hx^d)[k,:]\|\cdot \|\alpha\| \\
            =& \|AH_L(\hx^d)[k-1,:]-AH_L(x^d)[k-1,:]\|\cdot \|\alpha\| \\
            &+ \|H_L(x^d)[k,:]-H_L(\hx^d)[k,:]\|\cdot \|\alpha\| \\
            \leq& (\rho\|\hx^d-x^d\|+\|\hx^d-x^d\|)\|\alpha\| \\
            =&(1+\rho)\delta_d\|\alpha\|
        \end{split}
    \end{equation}
Notice that
    \begin{equation}\label{eq: bbw-bw}
    \begin{split}
        &\|\bbw_k^{(t)}-\bw_{t-L+k}\|=\|A(\tbw_{k-1}^{(t)}-\bw_{t-L+k-1})\| \\
        \leq &\rho\|\tbw^{(t)}_{k-1}-\bw_{t-L+k-1}\|
    \end{split}
    \end{equation}
Combine \eqref{eq: tbw-bbw} and \eqref{eq: bbw-bw} and we get
    \begin{equation}\label{eq: tbw-bw}
        \begin{split}
            &\|\tbw_k^{(t)}-\bw_{t-L+k}\| \\
            \leq &\|\tbw_k^{(t)}-\bbw_k^{t}\|+\|\bbw_k^{(t)}-\bw_{t-L+k}\| \\
            \leq& (1+\rho)\delta_d\|\alpha\|+ \rho\|\tbw_{k-1}^{(t)}-\bw_{t-L+k-1}\|
        \end{split}
    \end{equation}
Since $\|\hbw_{t}-\bw_t\|=\|\tbw_{L}^{(t)}-\bw_{t-L+L}\|$ and $\|\hbw_{t-L+1}-\bw_{t-L+1}\|\leq\beta$, by \eqref{eq: tbw-bw},
    \begin{equation}\label{eq: hbw-bw, with alpha}
        \|\hbw_{t}-\bw_t\|\leq (\beta-\frac{(1+\rho)\|\alpha\|\delta_d}{1-\rho})\rho^L+\frac{(1+\rho)\|\alpha\|\delta_d}{1-\rho}
    \end{equation}

    To further bound the RHS of \eqref{eq: tbw-bw}, we need to bound $\|\alpha\|$. Notice that $\frac{\|H_{ux}\|}{\|H_{ux}H_{ux}^T\|}=\frac{1}{\|H_{ux}\|}$, then by \eqref{eq: psudo inv alpha} and \eqref{eq: bound Hux} we can bound $\|\alpha\|$ as
    \begin{equation}\label{eq: bound alpha}
        \begin{split}
            \|\alpha\|\leq& \frac{1}{\|H_{ux}\|}(\|u_{t-L+2:t+1}\|+\|x_{t-L+2}-\hbw_{t-L+1}\|) \\
            \leq& LD\sqrt{L-1}(\beta+\frac{\varepsilon}{1-\rho})+\gamma+\beta+\frac{\varepsilon}{1-\rho} \\
        \end{split}
    \end{equation}

    Combine \eqref{eq: hbw-bw, with alpha} and \eqref{eq: bound alpha}, then we get 
    \begin{equation}\label{eq: hatw'-w'}
       \begin{split}
            &\|\hbw_t-\bw_t\| \\
            \leq &\rho^L\beta+\frac{1+\rho}{1-\rho^L}\delta_d((LD\sqrt{L-1}+1)(\beta+\frac{\varepsilon}{1-\rho})+\gamma)
       \end{split}
    \end{equation}
    Let $\delta_d\leq C_3$, $\beta=C_4\delta_d$ and $\gamma=\frac{\|B\|LD}{1-\rho}C_4\delta_d+C_5$ where $C_3,C_4,C_5$ are defined in Lemma \ref{lemma: w'-hatw'}. Then it can be verified that both \eqref{eq: |wt| initial} and \eqref{eq: x_t+1} hold, and RHS of \eqref{eq: hatw'-w'}$\leq\beta$. This completes the proof.
\end{itemize}

\section{Proof and Analysis for Section \ref{subsec: regret decom}}
\subsection{Proof of Lemma \ref{lemma: r1}}

Define 
$$\begin{cases}
    \tilde{\bM}^{*}=\arg\min_{\bM\in\mathbb{M}}\sum_{t=0}^{T-T_s-1}c_t(\bM,\hH_1,\hH_2,\{\hbw_t\})\\
    \tilde{f}_t(\bM_t)=c_t(\bM_t,\hH_1,\hH_2,\{\hbw_t\}) \\
    \tilde{\bu}_{t,L-1}=u_{t-L+2:t}(\bM_t|\{\hbw_t\})
\end{cases}$$
then
\begin{equation}\label{eq: r11+r12}
    \begin{split}
         R_1=&\sum_{t=0}^{T-T_s-1}{c_t(u_t,x_t)}-\sum_{t=0}^{T-T_s-1}c_t(u_t(\starM,\hH_1,\hH_2,\{\hbw_t\}) \\
            \leq&\sum_{t=0}^{T-T_s-1}{c_t(u_t,x_t)}-\sum_{t=0}^{T-T_s-1}c_t(u_t(\tilde{\bM}^{*},\hH_1,\hH_2,\{\hbw_t\})\\
            \leq & 
            \begin{Vmatrix}
                 \sum_{t=0}^{T-T_s-1}{c_t(u_t,x_t)}-{\sum_{t=0}^{T-T_s-1}{\tilde{f}_t(\bM_t)}}
            \end{Vmatrix} \\
            &+\begin{Vmatrix}
                 \sum_{t=0}^{T-T_s-1}{\tilde{f}_t(\bM_t)}-{\sum_{t=0}^{T-T_s-1}{\tilde{f}_t(\tilde{\bM}^{*})}}
            \end{Vmatrix}
    \end{split}
\end{equation}
By Lemma \ref{lemma: L-step} and Lemma \ref{lemma: simulation}, it follows that for $x_{t+1}$ and $x_t(\bM_t|\hH_1,\hH_2,\{\hbw_t\})$
\begin{subequations}\label{eq: x_t+1=}
    \begin{align*}
        x_{t+1}&=\sum_{k=1}^{[(t+1)/(L-1)]} H_2^{k-1}H_1 \bu_{t-(k-1)(L-1),L-1} + \bw_t \\
        &= \sum_{k=1}^{[(t+1)/(L-1)]} \hH_2^{k-1}\hH_1 \bu_{t-(k-1)(L-1),L-1} + \hbw_t
    \end{align*} 
\end{subequations}
and
\begin{equation*}\label{eq: tx_t+1=}
\begin{split}
    &x_{t+1}(\bM_t|\hH_1,\hH_2,\{\hbw_t\}) \\
    =&\sum_{k=1}^{[(t+1)/(L-1)]} \hH_2^{k-1}\hH_1 \bu_{t-(k-1)(L-1),L-1}(\bM_t|\{\hbw_t\}) + \hbw_t
\end{split}
\end{equation*}

Also, by Lemma \ref{lemma: w'-hatw'} and the property of OGD, we get that for $\forall i\leq t$
\begin{subequations}
     \begin{align*}
        &\|u_{t-i}-u_{t-i}(\bM_t|\{\hbw_t\})\| \\
        \leq &\sum_{j=1}^L\|M_{t-i}^{(j)}-M_t^{(j)}\|\cdot \|\hbw_{t-i-j}\| \\
        \leq & (\frac{\varepsilon}{1-\rho}+C_4\delta_d)L\lambda Gi\\
        =&(\frac{\varepsilon}{1-\rho}+C_4\delta_d)\frac{2L^2D}{\sqrt{T-T_s}} i \\
    \end{align*}
\end{subequations}
which implies
\begin{equation}\label{eq: u-tu}
\begin{split}
     &\|\bu_{t-(k-1)(L-1),L-1}-\tilde{\bu}_{t-(k-1)(L-1),L-1}\|\\
     \leq &\sum_{i=(k-2)(L-1)+1}^{(k-1)(L-1)}(\frac{\varepsilon}{1-\rho}+C_4\delta_d)\frac{2L^2D}{\sqrt{T-T_s}} i
\end{split}
\end{equation}

Define $\varrho=(\frac{1+\rho^L}{2})^{\frac{1}{L}}$. By \eqref{eq: Ts satisfies}and Lemma \ref{lemma: H-hatH} we have $C_H\delta_d\leq \varrho^L-\rho^L$. Combine this with \eqref{eq: x_t+1=},\eqref{eq: tx_t+1=} and \eqref{eq: u-tu}, we have
\begin{equation}\label{eq: r11}
    \begin{split}
        &\|x_t-x_t(M_t|\hH_1,\hH_2,\{\hbw_t\})\|\\
        \leq &\sum_{k=1}^{[(t+1)/(L-1)]}\|\hH_2^{k-1}\hH_1\|\cdot \|\bu_{t-(k-1)(L-1),L-1}\\
        &-\tilde{\bu}_{t-(k-1)(L-1),L-1}\| \\
        \leq & \sum_{k=1}^{[(t+1)/(L-1)]} \|\hH_1\|\|\hH_2\|^{k-1}(\frac{\varepsilon}{1-\rho}+C_4\delta_d)\cdot \\
        &\sum_{i=(k-2)(L-1)+1}^{(k-1)(L-1)}\frac{2L^2D}{\sqrt{T-T_s}} i \\
        \leq &(\frac{\|B\|}{1-\rho}+\delta_H)(\frac{\varepsilon}{1-\rho}+C_4\delta_d)\frac{2L^2D}{\sqrt{T-T_s}}\sum_{j=1}^t j\varrho^j \\
        \leq &\frac{2L^2D\varrho}{(1-\varrho)^2\sqrt{T-T_s}}(\frac{\|B\|}{1-\rho}+\delta_H)(\frac{\varepsilon}{1-\rho}+C_4\delta_d)
    \end{split}
\end{equation}

On the other hand, by the properties of OGD, we know that
\begin{equation}\label{eq: r12}
    \begin{split}
        &\begin{Vmatrix}
                 \sum_{t=0}^{T-T_s-1}{\tilde{f}_t(\bM_t)}-{\sum_{t=0}^{T-T_s-1}{\tilde{f}_t(\tilde{\bM}^{*})}}
            \end{Vmatrix}\\
            &\leq 2LDG\sqrt{T-T_s}
    \end{split}
\end{equation}

Applying \eqref{eq: r11} and \eqref{eq: r12} to \eqref{eq: r11+r12} and in view of $u_t(\bM_t|\{\hbw_t\}=u_t$, we have
\begin{equation}
    \begin{split}
        &R_1\leq  \begin{Vmatrix}
                 \sum_{t=0}^{T-T_s-1}{c_t(u_t,x_t)}-{\sum_{t=0}^{T-T_s-1}{\tilde{f}_t(\bM_t)}}
            \end{Vmatrix}\\
            &~~~~+\begin{Vmatrix}
                 \sum_{t=0}^{T-T_s-1}{\tilde{f}_t(\bM_t)}-{\sum_{t=0}^{T-T_s-1}{\tilde{f}_t(\tilde{\bM}^{*})}}
            \end{Vmatrix} \\
            &\leq G\sum_{t=0}^{T-T_s-1}\frac{2L^2D\varrho}{(1-\varrho)^2\sqrt{T-T_s}}(\frac{\|B\|}{1-\rho}+\delta_H)(\frac{\varepsilon}{1-\rho}+C_4\delta_d)\\
            &~~~~+ 2LDG\sqrt{T-T_s} \\
            &\leq 2LDG\bigl( \frac{L\varrho}{(1-\varrho)^2}(\frac{\|B\|}{1-\rho}+\delta_H)(\frac{\varepsilon}{1-\rho}+C_4\delta_d)\bigr)\sqrt{T-T_s} \\
            &\leq 2LDG\bigl( \frac{L\varrho}{(1-\varrho)^2}(\frac{\|B\|}{1-\rho}+C_H)(\frac{\varepsilon}{1-\rho}+C_4)\bigr)\sqrt{T-T_s}
    \end{split}
\end{equation}
This completes the proof of Lemma \ref{lemma: r1}.

\subsection{Proof of Lemma \ref{lemma: r2}}
Given a certain $\bM\in\mathbb{M}$, define $\hat{\bu}_{t,L-1}=u_{t-L+2:t}(\bM|\{\hbw_t\})$ and $\bar{\bu}_{t,L-1}=u_{t-L+2:t}(\bM|\{\bw_t\})$, by Lemma \ref{lemma: L-step} and Lemma \ref{lemma: simulation}, it holds for $x_{t+1}(\bM|H_1,H_2,\{\bw_t\})$ and $x_t(\bM|\hH_1,\hH_2,\{\hbw_t\})$ that
\begin{equation}\label{eq: xM and hxM}
    \begin{cases}
        x_{t+1}(\bM|H_1,H_2,\{\bw_t\}) \\
        =\sum_{k=1}^{[(t+1)/(L-1)]} H_2^{k-1}H_1 \bar{\bu}_{t-(k-1)(L-1),L-1} + \bw_t \\
        x_{t+1}(\bM|\hH_1,\hH_2,\{\hbw_t\}) \\
        =\sum_{k=1}^{[(t+1)/(L-1)]} \hH_2^{k-1}\hH_1 \hat{\bu}_{t-(k-1)(L-1),L-1} + \hbw_t
    \end{cases}
\end{equation}
We next quantify $\|\hat{\bu}\|$, $\|\hat{\bu}_{t,L-1}-\bar{\bu}_{t,L-1}\|$ and $\|H_2^{k}H_1-\hH_2^k\hH_1\|$ for any $k\in\mathbb{N}_+$. 

\begin{itemize}
    \item Quantify $\|\hat{\bu}\|$ and $\|\hat{\bu}_{t,L-1}-\bar{\bu}_{t,L-1}\|$: By Lemma \ref{lemma: w'-hatw'}, we have $\forall t=0,1,\dots, T-T_s-1$,
\begin{subequations}\label{eq: hatuM-baruM}
    \begin{align}
        \|\hat{\bu}_{t,L-1}\|&\leq \sum_{\tau=t-L+2}^t\sum_{i=1}^L\|M_\tau^{(i)}\|\|\hbw_{t-i}\| \\
        &\leq  L(L-1)D(C_4\delta_d+\frac{\varepsilon}{1-\rho})
    \end{align}
\end{subequations}
and also
\begin{equation}\label{eq: baruM}
    \begin{split}
        &\|\hat{\bu}_{t,L-1}-\bar{\bu}_{t,L-1}\| \\
        \leq &\sum_{\tau=t-L+2}^t\sum_{i=1}^L\|M_\tau^{(i)}\|\|\hbw_{t-i}-\bw_{t-i}\| \\
        \leq &L(L-1)DC_4\delta_d
    \end{split}
\end{equation}
\end{itemize}

\begin{itemize}
    \item Quantify $\|H_2^{k}H_1-\hH_2^k\hH_1\|$: By Lemma \ref{lemma: H-hatH} and the condition that $\delta_d\leq \frac{\varrho^L-\rho^L}{C_H}$, we have
    \begin{subequations}
        \begin{align}
            &\|H_2^{k}H_1-\hH_2^k\hH_1\|\\
            \leq &\|H_2^{k}H_1-H_2^k\hH_1\|+\|H_2^{k}\hH_1-\hH_2^k\hH_1\| \\
            \leq & \|H_2^k\|\|H_1-\hH_1\|+\|\hH_1\|\|H_2^k-\hH_2^k\| \\
            \leq &C_H\delta_d\rho^{Lk}+(\frac{\|B\|}{1-\rho}+C_H\delta_1)\|H_2^k-\hH_2^k\|
        \end{align}
    \end{subequations}
    Define $\varrho=(\frac{1+\rho^L}{2})^{\frac{1}{L}}$. By \eqref{eq: Ts satisfies} and Lemma \ref{lemma: H-hatH} we have $C_H\delta_d\leq \varrho^L-\rho^L$. By \cite[Lemma 17]{hazan2020nonstochastic},
    $$\|H_2^k-\hH_2^k\|\leq k\varrho^{Lk}\|H_2-\hH_2\|$$
    It follows that
    \begin{equation}\label{eq: H^k-hH^k}
    \begin{split}
        &\|H_2^{k}H_1-\hH_2^k\hH_1\|\\
        \leq &C_H\delta_d\rho^{Lk}+(\frac{\|B\|}{1-\rho}+C_H\delta_d)C_H\delta_dk\varrho^{Lk}
    \end{split}
    \end{equation}
\end{itemize}
Combining \eqref{eq: hatuM-baruM}, \eqref{eq: baruM} with \eqref{eq: H^k-hH^k} yields
\begin{equation}\label{eq: x_clean-hatx}
    \begin{split}
        &\|x_{t+1}(\bM|H_1,H_2,\{\bw_t\})-x_{t+1}(\bM|\hH_1,\hH_2,\{\hbw_t\})\| \\
        &\leq \sum_{k=1}^{[(t+1)/(L-1)]} \|H_2^{k-1}H_1\|\|\hat{\bu}_{t-(k-1)(L-1),L-1}-\\
        &\bar{\bu}_{t-(k-1)(L-1),L-1}\| \\
        &+\sum_{k=1}^{[(t+1)/(L-1)]}\|H_2^{k}H_1-\hH_2^k\hH_1\|\|\hat{\bu}_{t-(k-1)(L-1),L-1}\| \\
        &+ \sum_{k=1}^{[(t+1)/(L-1)]} \|\bw_t-\hbw_t\| \\
        &\leq \frac{\|B\|}{1-\rho}\sum_{k=0}^tLDC_4\delta_d\rho^k+LD(C_4\delta_d+\frac{\varepsilon}{1-\rho})\sum_{k=0}^t  C_H\delta_d\rho^{k} \\
        &+LD(C_4\delta_d+\frac{\varepsilon}{1-\rho})\sum_{k=0}^t(\frac{\|B\|}{1-\rho}+C_H\delta_d)C_H\delta_dk\varrho^{k}+C_4\delta_d \\
        &=\biggl( \frac{\|B\|LDC_4}{(1-\rho)^2}+ \frac{LDC_H\varepsilon}{(1-\rho)^2}\bigl(1+\frac{\|B\|}{(1-\varrho)^2}\bigr)+C_4 \biggr)\delta_d+o(\delta_d) \\
    \end{split}
\end{equation}
Apply \eqref{eq: x_clean-hatx}, \eqref{eq: hatuM-baruM} and the definition of $C_8,C_7$ to $R_2$ and we get
\begin{subequations}
    \begin{align*}
        &R_2=\sum_{t=0}^{T-T_s-1}c_t(\starM,\hH_1,\hH_2,\{\hbw_t\})-c_t(\starM,H_1,H_2,\{\bw_t\}) \\
        &\leq G\sum_{t=0}^{T-T_s-1}\|x_t(\bM^{*}|H_1,H_2,\{\bw_t\})-x_t(\bM^{*}|\hH_1,\hH_2,\{\hbw_t\})\|\\
        &~~~~+G\sum_{t=0}^{T-T_s-1}\|u_t(\starM|\{\hbw_t\})-u_t(\starM|\{\bw_t\})\| \\
        &\leq (GC_8+GLDC_4)\delta_d(T-T_s) + o(\delta_d)(T-T_s)
    \end{align*}
\end{subequations}
This completes the proof.

\newpage
\section{Algorithm with Output Feedback}\label{subsec: alg yt}
\begin{algorithm}
 	\caption{Data-Driven Online Adaptive Control Policy with Output Feedback ($\mathcal{A}_y$)}
        \label{alg: yt}
 	\begin{algorithmic}[1]
 	\Inputs{Time horizon $T$, numbers $I_0$, $L$, $N$, dimension $m$, $n$, $p$ set $\mathbb{M}$, gradient bound $G$.}
    \State \textbf{\textit{Stage 1}: Online exploration.}
 	\State Initialization: $t=0$, $x_0=0$,
 	\For{$k=0,1,\dots,I_0$}
            \For{$t=0,1,\dots,N$}
               \State Set $u_t^k=\{\pm1\}^{m}_{i.i.d}$ and collect $u_t^k$, $y_{t+1}^k$.
            \EndFor	
        \EndFor
        \State Build matrix $Y$ and $U$ as shown in Section 4. 
        \State Calculate $\hat{\Phi}_y=\frac{1}{I_0}YU^T$.
        \State Set $u^d=\{u_0^d,\dots,u_{N-1}^d\}$, $u_t^d\sim\text{Unif}\left(\frac{1}{\sqrt{N}}\mathcal{S}^{m-1}\right)$
        \State $\hy^d=\hPhi_y u^d$. 
        \State Build Hankel matrices $H(\hy^d)$ and $H(u^d)$. 
	\State \textbf{\textit{Stage 2}: Commitment in Noisy Environment.}
        \State Initialization: $t=0$, $T_s=NI_0$, $x_{t\leq 0}=0$, $u_{-L\leq t\leq T-T_s}=0$, $\hbw_{t\leq 0}=0$,  $M_{i\leq 0}^{(j)}=0$,$\forall 1\leq j \leq L$, $\lambda=\frac{2LD}{G\sqrt{T}}$.
       \For{$t=0,\dots,T-T_s$}
		\State Set $u_t=\sum_{i=1}^L M_t^{(i)}\hbw^{o}_{t-i}$. 
		\State Receive $y_{t+1}$ and $c_t(u_t,y_t)$. 
		\State Calculate $\hbw_t=\text{AccNoise}$$(u_{t-L+2:t}$, $y_{t-L+2}$, $y_{t+1}$, $\hbw_{t-L+1}^{o}$, $H_L(u^{d})$, $H_L(\hy^d))$. 
		\State Calculate $\ty_t(\bM_t)=\text{PiTraj}$$(\hbw_{0:t-1}^{o}$,$\bM_t$, $H_L(u^{d})$, $H_L(\hy^d),t)$. 
        \State $f_t(\bM_t)=c_t(u_t,\ty(\bM_t))$. 
		\State OGD: $\bM_{t+1}=\proj{M}{\bM_t-\lambda\nabla f_t(\bM_t)}$.
        \EndFor
    \end{algorithmic}
 \end{algorithm}
\end{document}